\documentclass[twocolumn, twocolappendix]{aastex631}

\usepackage{booktabs}
\usepackage{amsmath, multirow}
\usepackage{rotating}
\usepackage{longtable}

\usepackage[normalem]{ulem}     % striking through
\usepackage{soul,xcolor}        % colored strikethrough line
\setstcolor{red}                % set color of strikethough line
\usepackage{hyperref}           % hyperlink
\usepackage{threeparttable}     % footnote of table

\def\HI{{\rm H\,{\textsc{\romannumeral 1}}}}

\def\degr{{$\hbox{$^\circ$}$}}

\def\fdg{{$.\!\!^\circ$}}

\graphicspath{{./}{Figures/}}
\begin{document}

\title{FAST Ultra-Deep Survey: the baryonic Tully-Fisher relation in FUDS0 field}

\correspondingauthor{Hongwei Xi}
\email{hwxi@nao.cas.cn}
\correspondingauthor{Lister Staveley-Smith}
\email{lister.staveley-smith@uwa.edu.au}
\correspondingauthor{Bo Peng}
\email{pb@nao.cas.cn}

\author[0000-0001-6642-8307]{Hongwei Xi}
\affiliation{National Astronomical Observatories, Chinese Academy of Sciences\\
20A Datun Road, Chaoyang District, Beijing 100101, China}

\author[0000-0002-8057-0294]{Lister Staveley-Smith}
\affiliation{International Centre for Radio Astronomy Research (ICRAR), University of Western Australia,\\ 
35 Stirling Hwy, Crawley, WA 6009, Australia}

\author[0000-0001-6956-6553]{Bo Peng}
\affiliation{National Astronomical Observatories, Chinese Academy of Sciences\\
20A Datun Road, Chaoyang District, Beijing 100101, China}
\affiliation{Department of Astronomy and Institute of Interdisciplinary Studies, Hunan Normal University\\
Changsha, Hunan 410081, China}

\author[0000-0002-0196-5248]{Bi-Qing For}
\affiliation{International Centre for Radio Astronomy Research (ICRAR), University of Western Australia,\\ 
35 Stirling Hwy, Crawley, WA 6009, Australia}

\author[0000-0002-1311-8839]{Bin Liu}
\affiliation{National Astronomical Observatories, Chinese Academy of Sciences\\
20A Datun Road, Chaoyang District, Beijing 100101, China}

\author[0000-0002-7550-0187]{Dejian Ding}
\affiliation{National Astronomical Observatories, Chinese Academy of Sciences\\
20A Datun Road, Chaoyang District, Beijing 100101, China}
\affiliation{School of Astronomy and Space Science, University of Chinese Academy of Sciences\\
No.1 Yanqihu East Road, Huairou District, Beijing, 101408, China}

\author{Jianbin Li}
\affiliation{National Astronomical Observatories, Chinese Academy of Sciences\\
20A Datun Road, Chaoyang District, Beijing 100101, China}
\affiliation{School of Astronomy and Space Science, University of Chinese Academy of Sciences\\
No.1 Yanqihu East Road, Huairou District, Beijing, 101408, China}



\begin{abstract}

    The Baryonic Tully-Fisher relation (BTFR) is one of the tightest scaling relations for disk galaxies in the local Universe, and therefore is an important tool for studying the fomation and evolution of galaxies. However, the evolution of the BTFR over cosmic time is poorly understood due to the limited sample of \HI\ galaxies beyond the local Universe, limitations of optically-derived rotation curves, and selection effects. In this work, we explore the BTFR at redshifts up to $z=0.42$ from galaxies detected in the pilot FAST Ultra-Deep Survey (FUDS) field, FUDS0. As found in previous work, we identify two components in the plane of baryonic mass versus rotational velocity, $C_{\rm BTFR}$ (tight) and $C_{\rm Outlier}$ (dispersed). A Gaussian mixture model is employed to recover the BTFR, yielding the best fit parameters for the slope $k=3.32_{-0.11}^{+0.12}$, zero point $b=10.07_{-0.03}^{+0.03}$, and intrinsic scatter $\sigma_{\rm BTFR}=0.036_{-0.009}^{+0.010}$. A random forest classifier is used to investigate the origin of the outlier component. We find that low signal significance and inaccurate inclinations are the key factors that contribute to the outlier population, indicating that observational effects are the dominant origin. Evolutionary trends are examined in three different redshift bins. Both the slope and zero point show consistency within 1-$\sigma$ uncertainty in the two low redshift bins, indicating no significant evolution. The indirectly inferred BTFR parameters from the $C_{\rm Outlier}$ component in the highest redshift bin aligns with the conclusion. The ongoing full FUDS survey will provide a larger sample to enable more accurate constraints on BTFR evolution.
    
    % word limit: 250 words

\end{abstract}

\keywords{Extragalactic astronomy (506) --- Galaxy distances (590) --- Galaxy kinematics (602) --- HI line emission (690) --- Spiral galaxies (1560)}


\section{Introduction}

    For rotationally supported system, the Tully-Fisher relation (TFR; \citealp{1977A&A....54..661T}) is a tight scaling relation between luminosity and rotational velocity. However, the slope and intercept vary from band to band, preventing universal comparisons between different studies, even where a secondary parameter is introduced to minimize the intrinsic scatter \citep{2025A&A...696A..52B}. Alternative forms, such as stellar-mass TFR (STFR) and the baryonic TFR (BTFR), have somewhat lower scatter and therefore appear to be more fundamental in nature \citep{1999ASPC..170....3F, 2000ApJ...533L..99M}. In the $\Lambda$CDM model, a power-law relation can be derived using virial theorem by replacing $R_{\rm vir}$ in $\frac{GM_{\rm vir}}{R_{\rm vir}} = V_{\rm vir}^2$ with $M_{\rm vir} \propto \rho R_{\rm vir}^3$. This gives the relation $M_{\rm vir} \propto V_{\rm vir}^3$, with a slope of 3 in the log-log plane (similarly derived relations can be found in \citealp{1998MNRAS.295..319M, 1999ApJ...513..555S}). In spite of the complicated link between virial and observable quantities, these heuristic relations provide a reasonable match with observations. Detailed simulations of disk galaxy formation appear to confirm the validity of this approximate relation \citep[e.g.][]{2023MNRAS.520.3895G}. Since rotational velocity is not dependent of distance, the low scatter of the TFR in the local Universe means that distances can be measured independently of redshift and cosmological model. Hence, the TFR is commonly used to measure the local Hubble constant and deviations from uniform expansion \citep{2020AJ....160...71S, 2024MNRAS.533.1550B} and cosmic flow \citep{2008Ap.....51..336K, 2013AJ....146...86T, 2020MNRAS.498.2703B, 2020ApJ...902..145K}.

    The TFR has been been studied using strong optical emission lines, such as H$\alpha$ \citep{2012MNRAS.425..296M, 2011MNRAS.417.2347R, 2011MNRAS.416.1936T}, H$\beta$, [OII] \citep{2006A&A...455..107F} and [OIII], and at millimeter wavelengths with the CO line \citep{1992ApJ...393..530D, 1994A&A...283...21S, 2007MNRAS.377..806C, 2009ApJ...697..115C}. However, these lines are correlated with star formation, and tend to probe the inner part of galaxy rotation curves and can be prone to disturbance by increased velocity dispersion. On the contrary, the distribution of neutral hydrogen (\HI) normally extends to larger radii in galaxies, and better traces the dark matter halo. Hence, rotation velocities can be better measured using \HI\ lines.

    Interferometers have been intensively employed to map the \HI\ distribution in galaxies and measure velocities in the outer flat parts of their rotation curves. Such measurements have the added benefit of allowing independent kinematical measurements of disk inclination \citep{2008AJ....136.2648D, 2018MNRAS.478.1611K, 2021MNRAS.508.1195P, 2022PASA...39...59D}. The largest study of BTFR in the local Universe is based on the Spitzer Photometry \& Accurate Rotation Curves (SPARC), which includes 175 spiral galaxies with rotation curves measured from \HI\ \citep{2016AJ....152..157L}. New interferometer surveys such as the Widefield ASKAP $L$-band Legacy All-sky Blind surveY (WALLABY; \citealp{2020Ap&SS.365..118K}) are providing larger \HI\ samples and will further refine our understanding of the BTFR in the local Universe \citep{2023MNRAS.519.4589C, 2024MNRAS.533..925M, 2024ApJ...976..159D}.
    
    The SPARC comparison of different velocity definitions \citep{2019MNRAS.484.3267L}, including $V_{\rm flat}$, $V_{\rm max}$, allows  investigation of the most fundamental form of the TFR. Moreover, it enables a better comparison with velocity width definitions such as $W_{50}$ and $W_{20}$, typically measured in larger single-dish surveys such as the \HI\ Parkes All Sky Survey (HIPASS; \citealp{2001MNRAS.322..486B, 2008MNRAS.391.1712M}) and the Arecibo Legacy Fast ALFA Survey (ALFALFA; \citealp{2005AJ....130.2598G}). 
    
    Studies of BTFR beyond the local Universe allow better understanding of the evolution of galaxies. Predictions from simulations using semi-analytical \citep{2009ApJ...698.1467O} and hydrodynamical models \citep{2021MNRAS.507.3267G} suggest that at higher redshifts the BTFR will have a flatter slope, a higher intercept, and a larger scatter. Unfortunately, this is difficult to investigate owing to the weakness of the \HI\ 21\,cm emission line, so there are only limited studies at higher redshift. Based on 67 \HI\ galaxies from the MeerKAT International GigaHertz Tiered Extragalactic Exploration \HI\ survey (MIGHTEE-\HI; \citealp{2016mks..confE...6J}) at $z<0.081$, \citet{2021MNRAS.508.1195P} found no significant evolution in the BTFR. \citet{2023MNRAS.519.4279G} also found no evolution based on 36 \HI\ galaxies at $z \sim 0.2$ detected in the Blind Ultra Deep \HI\ Environmental Survey (BUDHIES; \citealp{2020MNRAS.496.3531G}). Similar conclusions were drawn from galaxies in the HIGHz survey \citep{2015MNRAS.446.3526C} using 39 \HI\ galaxies at similar redshifts. 
    
    At higher redshift, where \HI\ surveys currently are not available, optical observations of rotation velocities, combined with indirect estimates of gas mass, shed some light on the likely evolution of BTFR. However, no firm conclusions can yet be drawn. \citet{2017ApJ...842..121U} found an increase of zero point with redshift at $0.9<z<2.3$. \citet{2021A&A...647A.152A} did not find any evolution at $0.5<z<0.8$ compared to local results. \citet{2024A&A...689A.318S} discern a marginal evolution with a shallower slope for $0.6<z<2.5$.

    The FAST Ultra-Deep Survey (FUDS; \citealp{2022PASA...39...19X}) is an \HI\ survey aimed at detecting distant galaxies to explore gas evolution. In our pilot survey, \citet{2024ApJS..274...18X} discovered 128 \HI\ galaxies with redshifts up to $z\sim0.4$, which gives a wider redshift baseline for examining the evolution of TFR using \HI. This paper is composed as follows. In Section \ref{Sct_02}, we describe the data sets employed in the paper. The data processing procedures are introduced in Section \ref{Sct_03}. The properties of FUDS0 galaxies are derived in Section \ref{Sct_04}. Section \ref{Sct_05} details the methods employed for deriving the BTFR. The results are presented in Section \ref{Sct_06}, and further discussed in Section \ref{Sct_07}. Finally, we summarize our findings in Section \ref{Sct_08}. We employ a flat universe model with cosmological parameters of $\Omega_{\rm M}=0.3$, $\Omega_{\Lambda}=0.7$, and Hubble constant of $H_0 = 70\,h_{70}$\,km\,s$^{-1}$\,Mpc$^{-1}$ throughout the paper, unless mentioned otherwise.

\section{Data}\label{Sct_02}
    
    The Five-hundred-meter Aperture Spherical radio Telescope (FAST; \citealp{2011IJMPD..20..989N}) is the largest single dish radio telescope. Its large illuminated area, 19-beam feed \citep{8105012} and wide band receiver make it ideal for deep \HI\ surveys. Our FAST Ultra-Deep Survey (FUDS; \citealp{2022PASA...39...19X}) has therefore been targeting fields to detect faint, high redshift \HI\ galaxies to explore the evolution of cool gas in galaxies. We have concluded observations and data reduction for the pilot FUDS0 field. The corresponding FUDS0 catalog consists of 128 \HI\ galaxies with redshifts from 0 to 0.4 \citep{2024ApJS..274...18X}. We have also identified their multiwavelength counterparts \citep{2025ApJS..278...15X} in the ultraviolet (UV) bands from the Galaxy Evolution Explorer (GALEX GR6\footnote{\url{https://galex.stsci.edu/GR6/}}; \citealp{2005ApJ...619L...1M}), at optical wavelengths from the Sloan Digital Sky Survey (SDSS DR15\footnote{\url{https://skyserver.sdss.org/dr15/en/home.aspx}}; \citealp{2019ApJS..240...23A}), the Dark Energy Spectroscopic Instrument Legacy Imaging Survey (DESI LIS DR9\footnote{\url{https://www.legacysurvey.org/dr9/}}; \citealp{2019AJ....157..168D}), and in the Infrared (IR) bands from the unblurred Wide-field Infrared Survey Explorer (unWISE\footnote{\url{https://unwise.me/}}; \citealp{2019ApJS..240...30S}). Spectral Energy Distribution (SED) fits were performed on the broadband spectra to derive physical properties, such as stellar mass ($M_*$) and star formation rate (SFR).

    Source position, shape parameters and photometry were obtained from the DESI LIS data using \textsc{Tractor}\footnote{\url{https://github.com/dstndstn/tractor}}, which simultaneously fits images from all bands, assuming the same morphological model. The model is either a round exponential, a de Vaucouleurs profile, an exponential profile, or a S\'{e}rsic model. We adopt the shape parameters\footnote{The detail interpretation can be found in \url{https://www.legacysurvey.org/dr10/catalogs/}}, ellipticities (i.e. $shape\_e1$ ($\epsilon_1$) and $shape\_e2$ ($\epsilon_2$)) from DESI LIS DR10, to derive the axial ratio $\frac{b}{a}$, using the following equation.
    \begin{equation}
        \frac{b}{a} = \frac{1-|\epsilon|}{1+|\epsilon|} ,
        \label{Equ_01}
    \end{equation}
    where $\epsilon = \sqrt{\epsilon_1^2 + \epsilon_2^2}$. The 1-$\sigma$ uncertainty was estimated using a Monte Carlo (MC) method. The g-band value of $\frac{b}{a}$ from SDSS was also employed for comparison, but is based on shallower data than for DESI LIS.

    \subsection{Inclined Galaxy Sample (IGS)}

        \begin{figure*}
            \begin{center}
                \includegraphics[width=0.65\columnwidth]{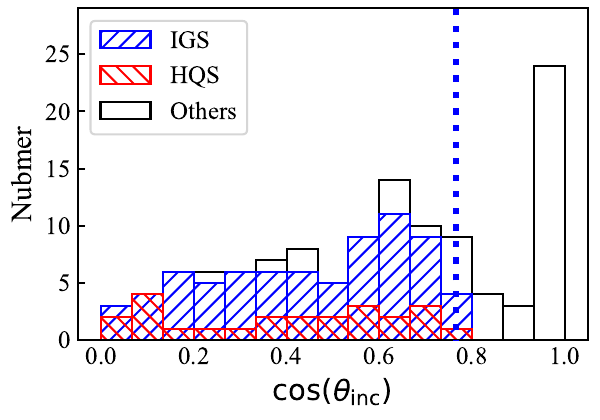}
                \includegraphics[width=0.65\columnwidth]{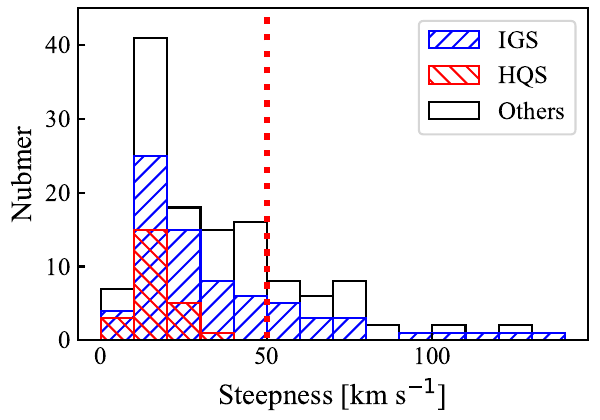}
                \includegraphics[width=0.65\columnwidth]{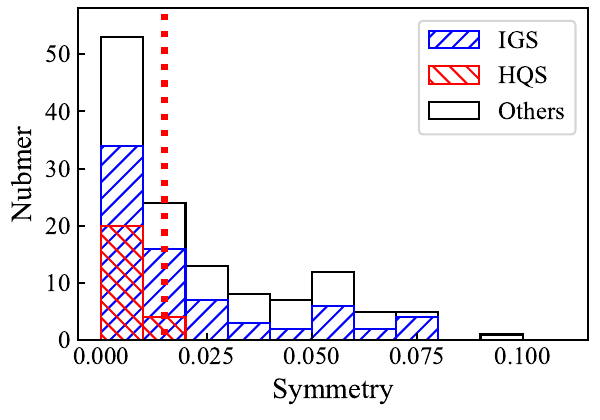}
                \includegraphics[width=0.65\columnwidth]{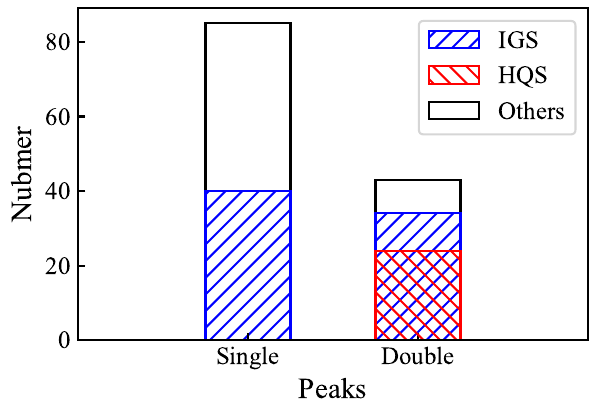}
                \includegraphics[width=0.65\columnwidth]{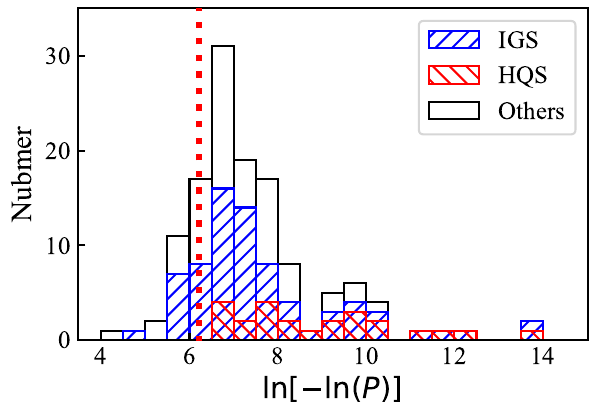}
                \includegraphics[width=0.65\columnwidth]{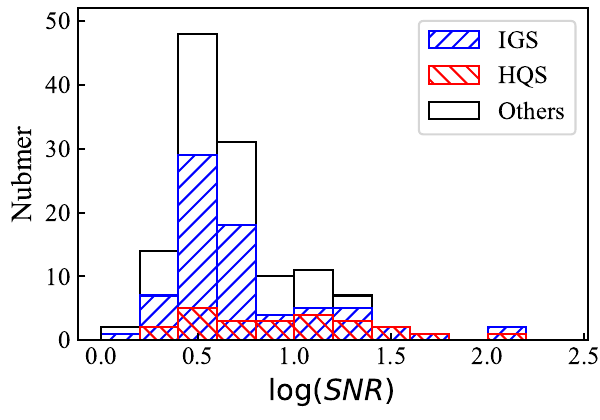}
                \includegraphics[width=0.65\columnwidth]{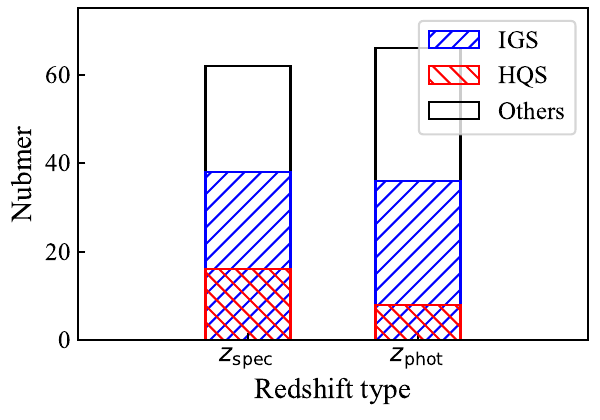}
                \includegraphics[width=0.65\columnwidth]{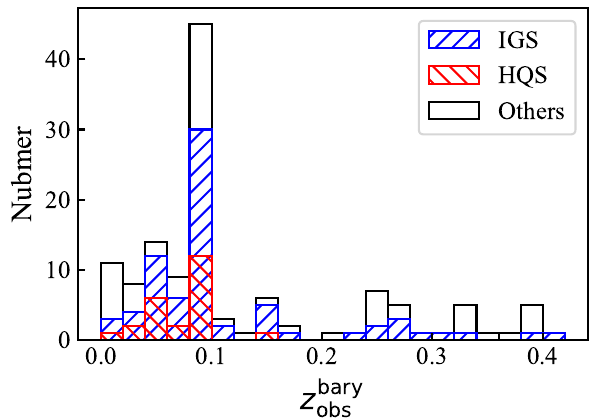}
                \includegraphics[width=0.65\columnwidth]{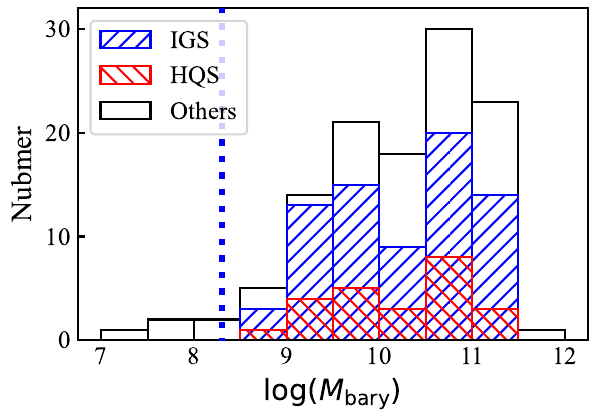}
                \caption{The distributions of inclination angle ($\cos(\theta_{\rm inc})$), steepness, symmetry, peak number, signal significance, $SNR$, optical redshift type, \HI\ redshift, and baryonic mass (from left to right, top to bottom). The blue (red) dotted lines indicate the cutoff values for IGS (HQS) sample. Note that the cutoff values for IGS are also applied on HQS (see text).}\label{Fig_01}
            \end{center}
        \end{figure*}

        We applied the following criteria to select galaxies suitable for studying the BTFR:
        \begin{enumerate}
        
            \item Stellar mass available from \citet{2025ApJS..278...15X}. 117 out of 128 \HI\ galaxies satisfy this criterion.
            
            \item Optical counterparts from DESI LIS lie within the FAST beam at the frequency of the \HI\ detection. 
            This helps to avoid poor estimates of the \HI\ flux. 
            This criterion reduces the sample to 113 \HI\ galaxies (all the galaxies satisfying first criterion have counterparts in DESI LIS). 
            
            \item Inclination angle (calculated in Section \ref{Sct_04_01_02}) larger than 40$^\circ$. This reduces uncertainty in the inclination correction to linewidth. This criterion results in a sample of 79 \HI\ galaxies.
            
            \item Baryonic mass (computed in Section \ref{Sct_04_02}) larger than $10^{8.3}\,h_{70}^{-2} {\rm M}_\odot$. The sample size is small at low baryonic masses, where the intrinsic scatter generally increases. This criterion further reduces the sample to 74 \HI\ galaxies.
            
        \end{enumerate}
        The final IGS sample consists of 74 \HI\ galaxies for further analysis. This sample has looser criteria than the following high-quality sample (HQS), but covers a wider redshift range.

    \subsection{High Quality Sample (HQS)}

        Significant scatter remains in rotational velocity versus baryonic mass plane for the above IGS sample. To better separate intrinsic and observational scatter, we therefore apply additional selection criteria to give our high quality sample (HQS). The extra criteria are: 
        \begin{enumerate}
            \addtocounter{enumi}{+4}
            
            \item \HI\ profile steepness $|W_{\rm m, 20}^{\rm rest}-W_{\rm m, 50}^{\rm rest}| < 50$ km\,s$^{-1}$ (as adopted by \citealp{2023MNRAS.519.4279G}). This reduces the sample to 58 \HI\ galaxies.

            \item \HI\ profile symmetry $|V_{\rm m, 20, cent}^{\rm rest} - V_{\rm m, 50, cent}^{\rm rest}|/W_{\rm m, 20}^{\rm  rest} < 0.015$, where $V_{\rm m, X, cent}^{\rm rest}$ is the center velocity of $W_{\rm m, X}^{\rm rest}$ in the rest frame (a larger value is adopted by \citealp{2023MNRAS.519.4279G}). This criterion results in a sample of 38 \HI\ galaxies.

            \item Double-horn \HI\ line profile, which was detected by code in best-fit spectra, $S_{\rm intr}(\nu)$. The sample is further reduced to 25 \HI\ galaxies.

            \item Signal significance $\ln(P) < -5 \times 10^2$, where $P$ is the likelihood of the line profile being drawn from a Gaussian distribution with a standard deviation equal to the local noise. This criterion removes both the strong narrow and weak wide galaxies and results in a final HQS sample of 24 HI galaxies.
            
        \end{enumerate}
        The 24-galaxy HQS sample has significantly less galaxies than our IGS sample, but has a much tighter relation in rotational velocity versus baryonic mass plane and is therefore a better reference sample for the true BTFR parameters, including intrinsic scatter.

        Figure \ref{Fig_01} shows the distribution of galaxies from the IGS and HQS samples, as well as the other galaxies in the sample (with available parameters), as a function of inclination angle, steepness, symmetry, number of peaks, signal significance, signal-to-noise ratio, optical redshift type, \HI\ redshift, and baryonic mass. The redshift ranges are $0.01<z<0.40$ in IGS, and $0.01<z<0.15$ in HQS. The HQS sample has a  narrower redshift coverage. The baryonic mass ranges are $8.61<\log(M_{\rm  bary}/h^{-2} {\rm M}_\odot)<11.47$ for IGS, and $8.94<\log(M_{\rm  bary}/h^{-2} {\rm M}_\odot)<11.47$ for HQS.

        In our previous work \citep{2024ApJS..274...18X}, the completeness of the catalog is parameterized as a function of noise-normalized mean \HI\ flux. In this paper, we employ the following equation to calculate the cumulative completeness ($C_{\rm cum}$) as a function of \HI\ mass for three samples (total sample, IGS, and HQS):
        \begin{equation}
            C_{\rm cum}(\log(M)) = \frac{\Sigma_i 1/V_{\rm com, tot}}{\Sigma_j 1/V_{{\rm com, det}, j}}
            \label{Equ_02}
        \end{equation}
        where $i$ and $j$ go through all galaxies with $\log(M_\HI)\geq \log(M)$ in the target sample (total sample, IGS, or HQS) and the FUDS0 catalog, respectively, $V_{\rm com, tot}$ is the total comoving volume in the FUDS0 survey, $V_{{\rm com, det}, j}$ is the detectable comoving volume for the $j$th galaxy. The detectable comoving volumes are computed by the method adopted in \citet{2015MNRAS.452.3726H} and \citet{2021MNRAS.501.4550X}.
        
\begin{figure}
            \begin{center}
                \includegraphics[width=0.95\columnwidth]{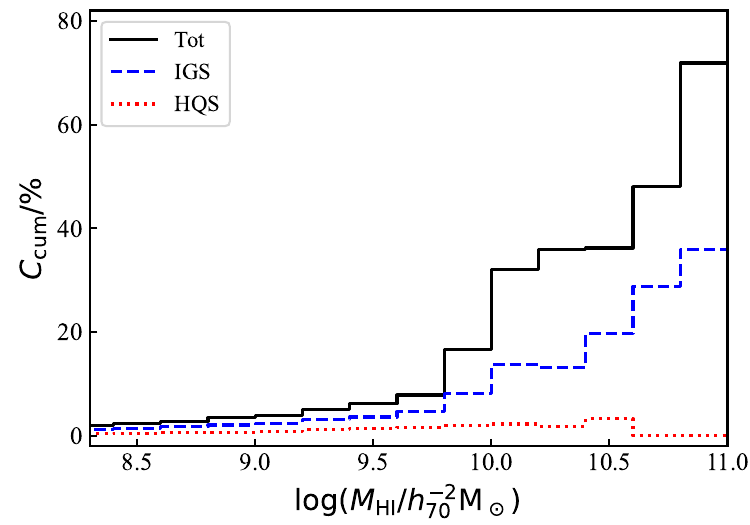}
                \caption{The cumulative completeness, $C_{\rm cum}$, as a function of \HI\ mass for the total (black solid), IGS (blue dashed), and HQS (red dotted) samples.}\label{Fig_02}
            \end{center}
        \end{figure}

        Figure \ref{Fig_02} displays $C_{\rm cum}$ for the total, IGS, and HQS samples. For the total sample, note that $C_{\rm cum}<10\%$ for low-mass and $\sim 70\%$ for high-mass galaxies. This is because FUDS0 surveys a large comoving volume beyond the local Universe, in which the low-mass galaxies are not detectable even with its high sensitivity of $\sim50$ $\mu$Jy beam$^{-1}$. Additionally, the sensitivity in the field gradually increases towards the edges, resulting in $C_{\rm cum}$ being less than 100\% even for the most massive \HI\ galaxies. In the IGS sample, $C_{\rm cum}$ gradually increases from $\sim 0\%$ to $\sim 35\%$. HQS has an extremely low $C_{\rm cum}$ ($\le5\%$) due to the rigorous criteria applied. Since there are no galaxies with $\log(M_\HI/h_{70}^{-2} {\rm M_\odot})\ge10.6$ in HQS, there is a sudden drop of $C_{\rm cum}$ at this mass.

\section{Reprocessing}\label{Sct_03}

    \subsection{Intrinsic \HI\ spectra}\label{Sct_03_01}

        The raw spectra have a frequency resolution of 7.63 kHz, and were smoothed by a Hanning window (0., 0.25, 0.75, 1., 0.75, 0.25, 0.) to a lower resolution of 22.9 kHz to increase SNR in the final position-position-frequency cube. The corresponding velocity resolutions are 1.61 and 4.83 km s$^{-1}$, respectively, at $z=0$. The intrinsic spectra of \HI\ galaxies need a small correction for the effects of finite velocity resolution.

        In our previous work \citep{2024ApJS..274...18X}, we employed Busy Function \citep{2014MNRAS.438.1176W} to model the spatially-integrated spectra of FUDS0 galaxies. Here, we also employ the Busy Function to model the intrinsic spectra but before binning and smoothing. The following procedure is used:
        \begin{enumerate}

            \item the spatially-integrated spectrum was modeled ($S_{\rm obs, mod}(\nu)$) using the parameters of Busy Function from our previous work \citep{2024ApJS..274...18X};
            
            \item the intrinsic spectrum ($S_{\rm intr}(\nu)$) was modeled using the above parameters;

            \item an integration was performed on the intrinsic spectrum to generate the binned spectrum ($S_{\rm intr, bin}(\nu)$);

            \item the above Hanning window was used to generate the smoothed spectrum ($S_{\rm intr, bin, smo}(\nu)$);

            \item an adjustment was performed on the parameters of $S_{\rm intr}(\nu)$ to find the best fit to minimize $\chi^2$ between the $S_{\rm intr, bin, smo}(\nu)$ and $S_{\rm obs, mod}(\nu)$. Note that the $S_{\rm intr, bin, smo}(\nu)$ was re-generated in steps 3 and 4 for each adjustment.

            \item $S_{\rm intr}(\nu)$ was derived with the best fit Busy Function parameters.
                
        \end{enumerate}

        We accordingly updated the FUDS0 observational parameters, including integrated flux $S_{\rm int}^{\rm bary}$\footnote{In this paper, we use the the subscript to describe the symbol, and superscript to denote the reference frame.}, flux-density-weighted frequency $\nu_{\rm cent}^{\rm bary}$, linewidth at 50\% and 20\% of the mean flux density ($\Delta \nu_{\rm m, 20}^{\rm bary}$, $\Delta \nu_{\rm m, 50}^{\rm bary}$ in frequency, with the mean flux density calculated within the 90\% the range of frequencies excluding 5\% of the integrated flux at each end of the profile; see \citealp{2009AJ....138.1938C} for details) and central frequency ($\nu_{\rm m, 20, cen}^{\rm bary}$, $\nu_{\rm m, 50, cen}^{\rm bary}$). The $\Delta \nu_{\rm m, X}$\footnote{X indicates the percentile at which the linewidth is measured.} velocity widths are adopted for the BTFR because they are more stable than the linewidth measured with reference to the peak flux density ($\Delta \nu_{\rm p, X}$), especially for widths at the 50\% level for asymmetrical line profiles. \citet{2009AJ....138.1938C} also find $W_{\rm m, X}$ (in velocity) is more consistent when used to predict the rotation velocity in the outer plat part of the rotation curve. 

        The shifted noise spectrum method was employed to estimate the uncertainties (see \citealp{2024ApJS..274...18X} for details). In this paper, we shifted the noise spectrum by one channel at a time. The intrinsic spectrum was derived for each shifted noise spectrum. Note that we also introduce extra 10\% uncertainty in flux density to better account for uncertainties in the calibration procedure. The final observational parameters are the median, and their uncertainties are calculated using the 15.87\% and 84.13\% percentiles. Compared with the results from \citet{2024ApJS..274...18X}, the median differences are 37\,kHz and 64\,kHz for $\Delta \nu_{20}^{\rm bary}$ and $\Delta \nu_{50}^{\rm bary}$, respectively, with maximum differences of 221\,kHz and 1303\,kHz, respectively. The extreme maximum difference is G087 (FUDS0 ID), due to its asymmetrical line profile. Otherwise, the average differences are only a few channel widths.

        The frequency linewidth in the barycentric frame was converted into the velocity linewidth in the rest frame of the galaxy using \citep{2017PASA...34...52M}:
        \begin{equation}
            W_{\rm m, X}^{\rm rest} \equiv \Delta V_{\rm m, X}^{\rm rest} \simeq  \frac{c (1 + z_{\rm cen}^{\rm bary})}{\nu_{\HI}^{\rm rest}} \Delta \nu_{\rm m, X}^{\rm bary} ,
            \label{Equ_03}
        \end{equation}
        where the redshift is derived by $z_{\rm cen}^{\rm bary} = \frac{\nu_{\HI}^{\rm rest} - \nu_{\rm cen}^{\rm bary}}{\nu_{\rm cen}^{\rm bary}}$, and $\nu_{\HI}^{\rm rest}$ is the rest frequency of 21 cm emission line.

    \subsection{Distances}\label{Sct_03_02}

        For the purposes of deriving distances and peculiar velocities from redshifts and positions, we initially adopt the cosmological parameters from Cosmicflows-3 \citep{2016AJ....152...50T}, namely $\Omega_{\rm M, cf3}=0.27$,  $\Omega_{\rm \Lambda, cf3} = 0.73$ and $H_{\rm 0, cf3} = 75$ km s$^{-1}$ Mpc$^{-1}$.  Using the derived values discussed below, we then re-calculate distances using the cosmological parameters employed in this paper, $\Omega_{\rm M}=0.3$, $\Omega_{\Lambda}=0.7$, and Hubble constant of $H_0 = 70\,h_{70}$\,km\,s$^{-1}$\,Mpc$^{-1}$.

        The online tool developed by \citet{2020AJ....159...67K} for the conversion between observed recessional velocity and luminosity distance is  available\footnote{\url{https://edd.ifa.hawaii.edu/calculator/}} for three different distance ranges. The first calculator employs the non-linear Numerical Action Method (NAM; \citealp{2017ApJ...850..207S}) to reconstruct the distance-velocity relation within 38\,Mpc based on the Cosmicflows-2 velocity model \citep{2014Natur.513...71T}. The second calculator (CF3 hereafter) extends the distance-velocity relation to 200\,Mpc using the linear model from \citet{2019MNRAS.488.5438G} based on Cosmicflows-3 data \citep{2016AJ....152...50T}. The most recent calculator (CF4), is available for the conversion up to 500\,Mpc based on a smoothed distance-velocity relation using Wiener filter model \citep{2024NatAs...8.1610V}. Considering that peculiar velocities mainly modify the redshift-distance relations in the local Universe, we only employed the NAM and CF3 versions, using NAM for $cz^{\rm bary} \leq 2,400$ km s$^{-1}$, CF3 for the remaining galaxies with $cz^{\rm bary} \leq 15,000$ km s$^{-1}$, and assuming pure motion from Hubble flow for galaxies beyond $cz^{\rm bary} = 15,000$ km s$^{-1}$.

        In the NAM calculator, the input parameter is the velocity in Standard Galactic of Rest (GSR) frame, using \citep{2020AJ....159...67K}:
        \begin{equation}
            \begin{split}
                V^{\rm gsr} = & V^{\rm hel} + 11.1\cos(l)\cos(b) + 251\sin(l)\sin(b)\\
                 & + 7.25\sin(b) ,
            \end{split}
            \label{Equ_04}
        \end{equation}
        where $(l, b)$ are the Galactic coordinates of the target. The output parameter is the luminosity distance ($D_{\rm lum}^{\rm gsr}$)\footnote{$D_{\rm lum}^{\rm gsr} = D_{\rm com}(1 + z^{\rm gsr})$, where $D_{\rm com}$ is the comoving distance. The beaming effect is neglected in this form.}. In the CF3 calculator, The input parameter is the velocity relative to Local Sheet (LS), which can be derived using the following equation \citep{2008ApJ...676..184T}:
        \begin{equation}
            \begin{split}
                V^{\rm ls} = & V^{\rm hel} -26 \cos(l)\cos(b)\\
                & + 317 \cos(l)\sin(b) - 8 \sin(l) .
            \end{split}
            \label{Equ_05}
        \end{equation}
        The luminosity distance in LS reference frame ($D_{\rm lum}^{\rm ls}$) is also returned by this calculator.

        As indicated above, the resultant luminosity and comoving distance estimates were then recalculated using the cosmological parameters adopted in this paper. For further details, see Appendix \ref{Sct_A} and \ref{Sct_B}.
        
\section{Galaxy properties}\label{Sct_04}

    \subsection{Rotational velocity linewidths}

        \subsubsection{Turbulent motion}

            The linewidth from integrated spectra is the combination of rotational motion and random motion from turbulence. It is generally assumed that the combination is linear summation for giant galaxies, and quadrature summation for dwarf galaxies. An equation was introduced by \citet{1985ApJS...58...67T} to describe this:
            \begin{equation}
                \begin{split}
                    (W_{\rm rot, X, proj}^{\rm rest})^2 = & (W_{\rm m, X}^{\rm rest})^2\\
                    & + W_{\rm tur, X}^2[1-2e^{-(\frac{W_{\rm m, X}^{\rm rest}}{W_{\rm c, X}})^2}]\\
                    & -2 W_{\rm m, X}^{\rm rest} W_{\rm tur, X}[1-e^{-(\frac{W_{\rm m, X}^{\rm rest}}{W_{\rm c, X}})^2}] ,
                \end{split}
                \label{Equ_06}
            \end{equation}
            where $W_{\rm rot, X, proj}^{\rm rest}$ is the projected rotation velocity estimated, $W_{\rm tur, X}$ is the random velocity, and $W_{\rm c, X}$ is the characteristic velocity for transition.

            We adopted $W_{\rm c, 20} = 120$ km\,s$^{-1}$ and $W_{\rm c, 50} = 100$ km\,s$^{-1}$ from \citet{2001A&A...370..765V}. Values for the turbulent component, $W_{\rm tur, X}$, vary widely. We adopt the result of \citet{1996PhDT........36R} for the local values: 
            \begin{equation}
                \begin{array}{rl}
                    & W_{\rm tur, 20}(z=0) = 30 \pm 3~{\rm km\,s}^{-1},\\
                    & W_{\rm tur, 50}(z=0) = 18 \pm 3~{\rm km\,s}^{-1},\\
                \end{array}
                \label{Equ_07}
            \end{equation}
            derived from measurements of the flat rotational velocity from 28 galaxies. \citet{2001A&A...370..765V} also found similar turbulent velocity in an independent sample of 22 galaxies in Ursa Major. \citet{2019ApJ...880...48U} modeled the evolution of atomic+molecular gas velocity dispersion based on 175 star-forming disk galaxies with $0.6<z<2.6$. Their redshift dependent term is employed to approximate the turbulent velocity evolution in this paper:
            \begin{equation}
                W_{\rm tur, X}(z) = W_{\rm tur, X}(z=0) + m z,
                \label{Equ_08}
            \end{equation}
            where $m=11.0\pm2.0$\,km\,s$^{-1}$.
            
        \subsubsection{Inclination}\label{Sct_04_01_02}

            The DESI LIS provides shape parameters, $\epsilon_1$ and $\epsilon_2$, which are measured by fitting a S\'{e}rsic brightness profile in all bands. The ratio between minor and major axes ($b/a$) can be computed from these parameters, following which we employ the following formula to convert to inclination angle ($\theta_{\rm inc}$):
            \begin{equation}
                \cos(\theta_{\rm inc}) = \sqrt{\frac{(b/a)^2- {\rm q}_0^2}{1-{\rm q}_0^2}} ,
                \label{Equ_09}
            \end{equation}
            where ${\rm q}_0 = 0.2$ is the ratio of b/a for an edge-on galaxy, as discussed in \citet{2000ApJ...533..744T}. The rotational velocity linewidth is then inferred from:
            \begin{equation}
                W_{\rm rot, X}^{\rm rest} = \frac{W_{\rm rot, X, proj}^{\rm rest}}{\sin(\theta_{\rm inc})} ,
                \label{Equ_10}
            \end{equation}
            where $W_{\rm rot, X}^{\rm rest}$ is the flat rotational velocity linewidth.

    \subsection{Baryonic mass}\label{Sct_04_02}

        \subsubsection{\HI\ mass}

            The \HI\ mass is derived using \citep{2017PASA...34...52M}:
            \begin{equation}
                M_{\HI} = 49.7 S_{\rm int}^{\rm bary} (D_{\rm lum}^{\rm bary})^2 ,
                \label{Equ_11}
            \end{equation}
            where $S_{\rm int}^{\rm bary}$ is the integral flux obtained in Section~\ref{Sct_03_01}, and $D_{\rm lum}^{\rm bary}$ is the luminosity distance from Section~\ref{Sct_03_02}. In our previous work \citep{2024ApJS..274...18X}, we  calculated the \HI\ mass using the luminosity distance from pure Hubble flow. The difference is only significant for  nearby galaxies, with median of 10.5\% and maximum of 28.0\% within $cz^{\rm bary}<15,000$\,km\,s$^{-1}$ or $D_{\rm com}<200\,h_{70}^{-1}{\rm Mpc}$. For more distant galaxies, the difference is mainly attributed to the correction to solar motion in the CMB frame, therefore is quite small.
    
        \subsubsection{Stellar mass}

            We have already cross-matched FUDS0 galaxies with UV, optical, and infrared (IR) sources \citep{2025ApJS..278...15X} and derived stellar masses using SED fits from \textsc{ProSpect} \citep{2020MNRAS.495..905R}, adjusted for consistency with the GSWLC-2 catalog \citep{2018ApJ...859...11S}. In this paper, we further adjust the stellar masses to use the luminosity distances derived in Section \ref{Sct_03_02}. As for \HI\ masses, the difference is small, except for nearby galaxies.

        \subsubsection{Molecular mass}\label{Sct_04_02_03}

            Statistical studies \citep{2017ApJS..233...22S, 2018MNRAS.476..875C} show that the contribution of molecular gas could be as high as that from \HI\ for high-mass galaxies. The results from \citet{2025MNRAS.544..193J} also indicate the importance of molecular mass at high redshift due to Malmquist bias, which is also seen in this work (see Section \ref{Sct_07_02}). Unfortunately, we do not have direct measurement of molecular mass. Instead, the empirical formula from \citet{2018ApJ...853..179T} was employed to predict the molecular mass. \citet{2018ApJ...853..179T} studied molecular gas in 1444 star-forming galaxies with $0<z<4$. They measured helium and molecular gas mass, $M_{\rm Hel, Mol}$, as a function of redshift, stellar mass, star formation rate, and optical effective radius. From their scaling relations, we estimate $M_{\rm Hel, Mol}$ for 117 FUDS0 galaxies, whose physical properties are derived in \citet{2025ApJS..278...15X}. Note that $M_{\rm Hel, Mol}$ is available for all the galaxies in IGS and HQS. The molecular mass is then calculated using $M_{\rm Mol} = (1-0.36) \times M_{\rm Hel, Mol}$, where 0.36 is the contribution from helium adopted in \citet{2018ApJ...853..179T}.

            \begin{figure}
                \begin{center}
                    \includegraphics[width=0.95\columnwidth]{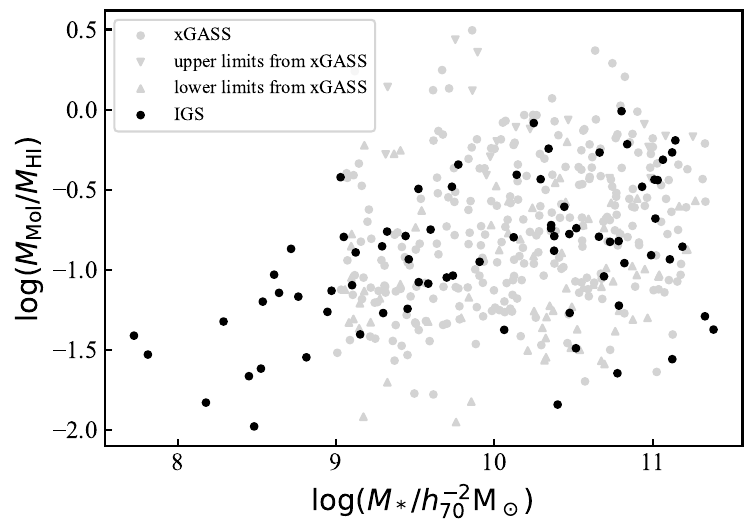}
                    \caption{The distribution of molecular fraction, $\log(M_{\rm Mol}/M_\HI)$, as a function of stellar mass. IGS galaxies are indicated by the black circles, while xGASS galaxies are indicated by gray dots, triangles (lower limits), and inverted triangles (upper limits).}\label{Fig_03}
                \end{center}
            \end{figure}

            Given this method depends on physical properties from optical data, we employed the independent \HI\ data to verify the predicted molecular mass. Figure \ref{Fig_03} shows the distribution of molecular fraction, $M_{\rm Mol}/M_\HI$, as a function of stellar mass. For comparison, we also show the galaxy distribution in the local Universe from the extended GALEX Arecibo SDSS Survey (xGASS, \citealp{2017ApJS..233...22S, 2018MNRAS.476..875C}). For consistency, we remove the 0.36 helium contribution from the xGASS data \citep{2017MNRAS.470.4750A}. Clearly, the IGS sample has similar distribution to xGASS sample. FUDS0 is an \HI-selected sample, therefore has more low stellar mass galaxies, while xGASS is stellar mass selected sample with $\log(M_*/h_{70}^{-2} {\rm M}_\odot)>9$. The consistency gives us confidence on the estimation of molecular mass.

            Finally, the total baryonic mass is derived from:
            \begin{equation}
                M_{\rm bary} = M_* + 1.4\times (M_{\HI} + M_{\rm Mol}),
                \label{Equ_12}
            \end{equation}
            where the factor of 1.4 accounts for the contribution by helium and metals in solar metallicity atomic gas.
            
\section{Methods}\label{Sct_05}

    \subsection{Models}

        \subsubsection{One-component model}

            The BTFR is a fundamental scaling relation, which holds across several orders of magnitude \citep{2000ApJ...533L..99M}. Usually, three parameters are used to describe the linear relation, slope ($k$), zero point ($b$), and intrinsic scatter ($\sigma_{\rm BTFR}$; perpendicular to the best fit line) \citep{2019MNRAS.484.3267L}. Here, we employ the following form to minimize the dependence of $b$ on $k$ in the fits:
            \begin{equation}
                y = k \times (x-2.4) + b .
                \label{Equ_13}
            \end{equation}
            In our HQS sample, the galaxies follow a tight power-law relation. Hence, we use a single Gaussian model to fit this sample.

            Taking into account the uncertainties of the data (rotation velocity and baryonic mass), the scatter of data along the direction perpendicular to the best fit line can be decomposed according to:
            \begin{equation}
                \sigma_\perp = \sqrt{\sigma_{\rm BTFR}^2 + \sigma_{\rm data, \perp}^2} ,
                \label{Equ_14}
            \end{equation}
            where $\sigma_{\rm data, \perp}$ is the observational uncertainty perpendicular to the line of best fit:
            \begin{equation}
                \sigma_{\rm data, \perp}^2 = [\sigma_x\sin(\theta_{\rm slope})]^2 + [\sigma_y\cos(\theta_{\rm slope})]^2 ,
                \label{Equ_15}
            \end{equation}
            with $\sigma_x$ ($\sigma_y$) being the uncertainty of the horizontal (vertical) data, and $\theta_{\rm slope}$ ($= \arctan(k)$) being the angle between the best fit line and the horizontal axis. Therefore, the data follows a Gaussian distribution with scatter of $\sigma_\perp$:
            \begin{equation}
                p(x, y, \sigma_x, \sigma_y| k, b, \sigma_{\rm BTFR}) = \frac{1}{\sqrt{2\pi} \sigma_\perp}\exp(-\frac{d_\perp^2}{2\sigma_\perp^2}) ,
                \label{Equ_16}
            \end{equation}
            where $d_\perp$ is the distance of the data point $(x, y)$ from the best fit line, as calculated from $d_\perp = |y-k \times (x-2.4)-b|/\sqrt{k^2 +1}$.

        \subsubsection{Two-component model}

            In our IGS sample, the galaxy distribution in the BTFR plane is better described using two Gaussian components. The first is the same as the single-component model above ($C_{\rm BTFR}$). The second is needed to better describe the outlier population ($C_{\rm Outlier}$). \citet{2023ApJ...950...87B} also demonstrated the need for a second component when they studied the BTFR using data from ALFALFA ($\alpha.R$). They attributed $C_{\rm Outlier}$ to radio frequency interference (RFI), confusion, and mismatched optical counterparts. They employed a Gaussian mixture model\footnote{\url{https://dfm.io/posts/mixture-models/}} \citep{foreman_mackey_2014_15856} to jointly fit the distribution. We use the same method for the IGS sample.

            Three parameters are used to describe the $C_{\rm BTFR}$ component: slope ($k$), zero point ($b$) and intrinsic scatter ($\sigma_{\rm BTFR}$). The $C_{\rm Outlier}$ component shares same slope as $C_{\rm BTFR}$, along with vertical offset from $C_{\rm BTFR}$ ($\delta_{\rm Outlier}$) and intrinsic scatter ($\sigma_{\rm Outlier}$). The fractional contribution of the $C_{\rm BTFR}$ component is described by an additional parameter, $f$. Similar to the one component model, the scatter including the data uncertainties can be expressed using:
            \begin{equation}
                \begin{array}{cc}
                     & \sigma_{\rm BTFR, \perp} = \sqrt{\sigma_{\rm BTFR}^2 + \sigma_{\rm data, \perp}^2}\\
                     & \sigma_{\rm Outlier, \perp} = \sqrt{\sigma_{\rm Outlier}^2 + \sigma_{\rm data, \perp}^2} ,
                \end{array}
                \label{Equ_17}
            \end{equation}
            where $\sigma_{\rm data, \perp}$ is given by Equation \ref{Equ_15}. The $C_{\rm BTFR}$ and $C_{\rm Outlier}$ components can be respectively described by:
            \begin{equation}
                \begin{array}{rl}
                    & p_{\rm BTFR}(x, y, \sigma_x, \sigma_y | k, b, \sigma_{\rm BTFR}) \\
                    = &\frac{1}{\sqrt{2\pi}\sigma_{\rm BTFR, \perp}}\exp(-\frac{d_{\rm BTFR, \perp}^2}{2\sigma_{\rm BTFR, \perp}^2}) \\
                    & p_{\rm Outlier}(x, y, \sigma_x, \sigma_y | k, b, \delta_{\rm Outlier}, \sigma_{\rm Outlier}) \\
                    = & \frac{1}{\sqrt{2\pi}\sigma_{\rm Outlier, \perp}}\exp(-\frac{d_{\rm Outlier, \perp}^2}{2\sigma_{\rm Outlier, \perp}^2}) ,
                \end{array}
                \label{Equ_18}
            \end{equation}
            where $d_{\rm BTFR, \perp}$ ($d_{\rm Outlier, \perp}$) is the distance of data point ($x, y$) from best fit line to $C_{\rm BTFR}$ ($C_{\rm Outlier}$), and can be derived from $d_{\rm BTFR, \perp} = |y-k \times (x-2.4)-b|/\sqrt{k^2 +1}$ ($d_{\rm Outlier, \perp} = |y-k \times (x-2.4) - b - \delta_{\rm Outlier}|/\sqrt{k^2 +1}$). The distribution of IGS galaxies can be modeled using the following equation:
            \begin{equation}
                \begin{array}{rl}
                     &  p(x, y, \sigma_x, \sigma_y | k, b, \sigma_{\rm BTFR}, \delta_{\rm Outlier}, \sigma_{\rm Outlier}, f) \\
                     = & f \times p_{\rm BTFR} + (1-f) \times p_{\rm Outlier} .\\
                \end{array}
                \label{Equ_19}
            \end{equation}

            For a data point with ($x, y, \sigma_x, \sigma_y$), we can use the model ($k, b, \sigma_{\rm BTFR}, \delta_{\rm Outlier}, \sigma_{\rm Outlier}, f$) to calculate the probability of the data point belonging to $C_{\rm BTFR}$:
            \begin{equation}
                prob = \frac{f \times p_{\rm BTFR}}{f \times p_{\rm BTFR} + (1-f) \times p_{\rm Outlier}} .
                \label{Equ_20}
            \end{equation}

    \subsection{Fitting}
        
        Here we employ the python package, \textsc{emcee}\footnote{\url{https://emcee.readthedocs.io/}} \citep{2013PASP..125..306F}, to implement Markov Chain Monte Carlo (MCMC) method to sample the posterior distribution of the model parameters. The best fit values are given by the median values, while their 1-$\sigma$ uncertainties are derived based on 15.9\% and 84.1\% percentiles. The likelihood function can be expressed as:
        \begin{equation}
            \ln(L) = \Sigma_i \ln(p_i) ,
            \label{Equ_21}
        \end{equation}
        where $p_i$ is the probability of galaxy $i$ in the model either from Equation \ref{Equ_16} for the one-component model or from Equation \ref{Equ_19} for the two-component model.

        We adopt a uniform distribution for each parameter as a prior. The limits of the distribution are given in Table \ref{Tab_01}. For the two-component model, we added an extra constraint ($\sigma_{\rm BTFR} \leq \sigma_{\rm Outlier}$) to avoid the obvious degeneracy between the two parameters. 50 Markov chains were generated to explore the posterior distribution in parameter space. We adopted the 1000 samples in each chain after removing 1000 burn-in steps. Then the procedures above were repeated once to get the final distribution with 2.25\% and 97.75\% percentiles (2-$\sigma$ uncertainty range) used to updated the limits, avoiding local minima caused by sample noise. The best fit values were derived based on the final distribution.

    \subsection{Verification}

        We performed a simulation to verify the viability of the methods adopted in this paper. The simulation produces mock IGS and HQS samples with ground-truth values for physical parameters (including $M_*$, $M_\HI$, $M_{\rm Mol}$, $M_{\rm bary}$, $W_{\rm rot}^{\rm rest}$) with known BTFR parameters. Their uncertainties are estimated using scaling relations based on the FUDS0 catalog. To mimic the deviations of the observed values from the ground-truth values, the observed values are resampled from a Gaussian distribution using the ground-truth values for the mean and the 1-$\sigma$ uncertainties. We applied the completeness of the FUDS0 catalog, the IGS selection criteria, and the redshift distribution to select the final sample based on the observed values. Note that the mock samples have the same galaxy numbers as the IGS and HQS samples. The details of the simulation are presented in Appendix \ref{Sct_C}.

        \begin{figure}
            \begin{center}
                \includegraphics[width=0.95\columnwidth]{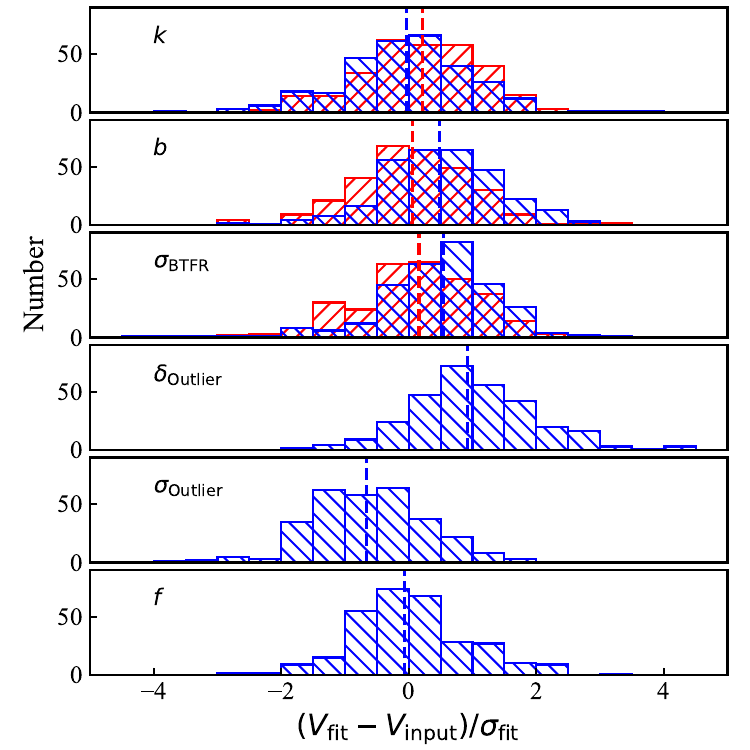}
                \caption{The distribution of normalized offsets of the parameters recovered by one-component (red) and two-component (blue) methods. The median values are indicated by dashed lines with the corresponding colors.}\label{Fig_04}
            \end{center}
        \end{figure}

        The one-component method was applied to observed values in the mock HQS sample to recover BTFR parameters. The normalized offset $\delta = \frac{V_{\rm fit} - V_{\rm input}}{\sigma_{\rm fit}}$ is employed to quantify the accuracy and precision of the results. The simulation and fitting are repeated 300 times to give the distribution of $\delta$. Figure \ref{Fig_04} shows the distribution of $\delta$. As a result, we found that all the normalized offsets, $\delta_k$, $\delta_b$, and $\delta_{\sigma_{\rm BTFR}}$, follow a Gaussian distribution with a standard deviation of $\sim 1$. The mean values are $\bar{\delta}_k= 0.22$, $\bar{\delta}_b=0.06$, and $\bar{\delta}_{\sigma_{\rm BTFR}}=0.16$, indicating only small systematic offsets.

        The same procedure was performed to apply the two-component method on the mock IGS sample to verify its validity. The distribution of $\delta$ is also given in Figure \ref{Fig_04}. Similar to the results from one-component method on the mock HQS, the normalized offsets of the parameters also follows a Gaussian distributions with standard deviation of $\sim 1$. The mean values are $\bar{\delta}_k=-0.04$, $\bar{\delta}_b=0.48$, $\bar{\delta}_{\sigma_{\rm BTFR}}=0.55$, $\bar{\delta}_{\delta_{\rm Outlier}}=0.92$, $\bar{\delta}_{\sigma_{\rm Outlier}}=-0.66$, and $\bar{\delta}_f=-0.06$. Compared with results from the one-component method, the BTFR parameters show larger systematic offsets, but are still within 1-$\sigma$. For the outlier parameters, the offsets are even larger, but, still within 1-$\sigma$. Since our analysis (see Section \ref{Sct_07_01}) confirms that the outlier component is caused by large measurement errors rather than a real scaling relation, and our interest is in the BTFR parameters, the poorly recovered outlier parameter ($\bar{\delta}_{\delta_{\rm Outlier}}$) does not impact the application of the method. Moreover, we are able accurately recover the BTFR fraction, $f$.

        \begin{figure}
            \begin{center}
                \includegraphics[width=0.95\columnwidth]{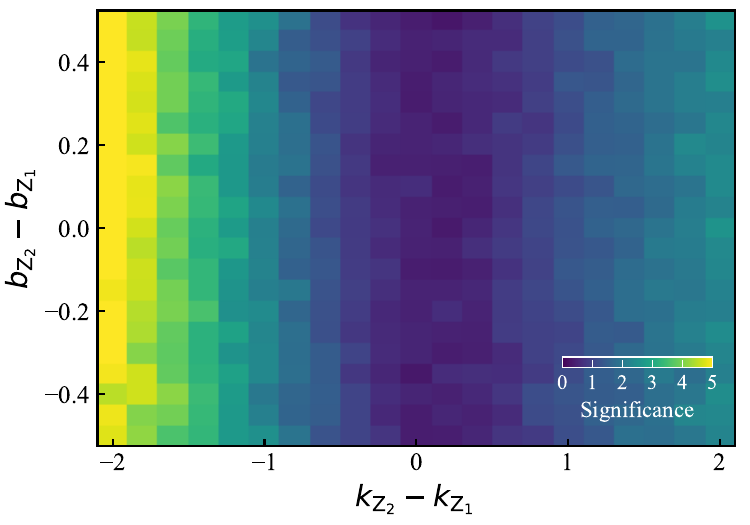}
                \includegraphics[width=0.95\columnwidth]{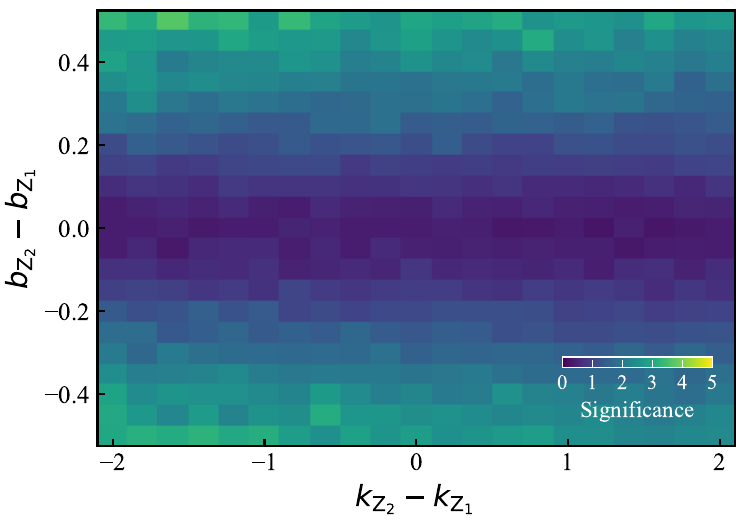}
                \caption{The significance of evolving $k$ (upper) and $b$ (lower) recovered by the two-component method as a function of the input parameter differences between the $z \leq 0.08$ and $0.08<z<0.12$ redshift bins.}\label{Fig_05}
            \end{center}
        \end{figure}

        The two-component method was also checked for its ability to recover the evolving BTFR. For simplicity, we employed a step function to simulate the evolution of BTFR parameters with fixed $k$ and $b$ (same values as those in the static BTFR) at low redshift ($z<0.08$), and varied $k$ and $b$ otherwise. The same procedure was employed to produce 50 mock catalogs for each pair of $k$ and $b$. The BTFR parameters were computed using the two-component method in three different redshift bins ($z_1$: $z\leq0.08$, $z_2$: $0.08<z\leq0.12$, and $z_3$: $z>0.12$, the same as that employed in Section \ref{Sct_07_02}). The median significance of recovered evolving $k$ and $b$ is shown as a function of the input parameter differences ($\Delta k$ and $\Delta b$) between $z_1$ and $z_2$ bins in Figure \ref{Fig_05}. A similar pattern with lower significance was discovered by comparing results in $z_1$ and $z_3$ redshift bins. The results indicate that the evolution of the BTFR slope and zero point can be identified by the two-component method if the evolution is strong enough. However, it is harder to detect in the higher redshift bin due to the narrower baryonic mass and linewidth span. In reality, the evolution of BTFR will differ from a simple step function, but the broad conclusions about sensitivity to evolution will remain similar.

        In summary, both the one-component and two-component methods can recover static BTFR parameters from mock data with measurement uncertainties and selection effects taken into account. Additionally, the two-component method is capable of measuring evolution of BTFR parameters, giving us confidence in the results recovered using the real data.

\section{Results}\label{Sct_06}

    Similar results were found when using either $W_{\rm m, 20}$ or $W_{\rm m, 50}$ as the proxy for rotational velocity. Hence, we only discuss $W_{\rm rot, 20}^{\rm rest}$ in the main text. Results for $W_{\rm rot, 50}^{\rm rest}$ are given in Appendix \ref{Sct_D}.

    \begin{table*}[!ht]
        \centering
        \caption{The best fit values in $M_{\rm bary}$ -- $W_{\rm rot, 20}^{\rm rest}$ plane. The limit ranges are given below the parameter symbols. The first row in each sample lists the results from the full sample. In IGS sample results, row 2 -- 4 give the best fit values in three redshift bins with slope free or fixed in fitting. In the highest redshift bin, we give extra results fitted by one-component model with $k$ and $b$ fixed in the last row for comparison. The last column lists the $b+\delta_{\rm Outlier}$ in each fitting, which indicates the zero point of $C_{\rm Outlier}$ component.}\label{Tab_01}
        \begin{tabular}{lrrrrrrrrr}
            \toprule
                &  & & \multicolumn{6}{c}{Parameters} \\
            \cmidrule(lr){4-10}
            Sample & $z$ & $N$ & $k$ & $b$ & $\sigma_{\rm BTFR}$ & $\delta_{\rm Outlier}$ & $\sigma_{\rm Outlier}$ & $f$ & $b+\delta_{\rm Outlier}$\\
              & & & [2, 5] & [2, 15] & [0.001, 0.25] & [0, 2] & [0.001, 0.5] & [0.01, 0.99] & - \\
            \midrule

            HQS & $0.00 - 0.42$ &24 & $3.38_{-0.17}^{+0.20}$ & $10.07_{-0.04}^{+0.04}$ & $0.045_{-0.007}^{+0.010}$ & - & - & - & - \\
            \cmidrule(lr){1-10}
            \multirow{8}{*}{IGS} & $0.00 - 0.42$ & 74 & $3.32_{-0.11}^{+0.12}$ & $10.07_{-0.03}^{+0.03}$ & $0.036_{-0.009}^{+0.010}$ & $0.47_{-0.13}^{+0.14}$ & $0.20_{-0.02}^{+0.03}$ & $0.57_{-0.09}^{+0.08}$ & $10.54_{-0.12}^{+0.14}$ \\
            \cmidrule(lr){2-10}
            & \multirow{2}{*}{$0.00 - 0.08$} & \multirow{2}{*}{25} & $3.50_{-0.22}^{+0.27}$ & $10.11_{-0.05}^{+0.06}$ & $0.039_{-0.013}^{+0.021}$ & $0.43_{-0.28}^{+0.51}$ & $0.27_{-0.08}^{+0.11}$ & $0.66_{-0.16}^{+0.13}$ & $10.52_{-0.27}^{+0.48}$\\
            &  & & $3.32$ [fix] & $10.09_{-0.08}^{+0.05}$ & $0.037_{-0.012}^{+0.019}$ & $0.45_{-0.29}^{+0.71}$ & $0.26_{-0.10}^{+0.12}$ & $0.64_{-0.26}^{+0.13}$ & $10.43_{-0.21}^{+0.41}$ \\
            \cmidrule(lr){2-10}
            & \multirow{2}{*}{$0.08 - 0.12$} & \multirow{2}{*}{32} & $3.06_{-0.23}^{+0.28}$ & $10.09_{-0.07}^{+0.06}$ & $0.043_{-0.016}^{+0.018}$ & $0.66_{-0.24}^{+0.24}$ & $0.14_{-0.04}^{+0.06}$ & $0.67_{-0.16}^{+0.14}$ & $10.74_{-0.23}^{+0.26}$\\
            &  & & $3.32$ [fix] & $10.06_{-0.05}^{+0.06}$ & $0.040_{-0.016}^{+0.019}$ & $0.67_{-0.24}^{+0.25}$ & $0.14_{-0.04}^{+0.06}$ & $0.64_{-0.16}^{+0.14}$ & $10.72_{-0.24}^{+0.27}$ \\
            \cmidrule(lr){2-10}
            & \multirow{3}{*}{$0.12 - 0.42$} & \multirow{3}{*}{17} & $3.97_{-1.00}^{+0.72}$ & $10.37_{-0.77}^{+0.36}$ & $0.163_{-0.067}^{+0.049}$ & $0.82_{-0.59}^{+0.67}$ & $0.26_{-0.07}^{+0.12}$ & $0.43_{-0.32}^{+0.41}$ & $10.97_{-0.33}^{+0.61}$\\
            &  & & $3.32$ [fix] & $10.46_{-0.68}^{+0.27}$ & $0.166_{-0.084}^{+0.050}$ & $0.69_{-0.50}^{+0.67}$ & $0.27_{-0.07}^{+0.12}$ & $0.53_{-0.39}^{+0.35}$ & $10.97_{-0.30}^{+0.53}$ \\
            &  & & $3.32$ [fix] & $10.21_{-0.19}^{+0.20}$ & - & $0.47$ [fix] & $0.23_{-0.04}^{+0.05}$ & - & $10.68_{-0.19}^{+0.20}$ \\
            
            \bottomrule
        \end{tabular}
    \end{table*}

    \subsection{HQS}

\begin{figure}
            \begin{center}
                \includegraphics[width=0.95\columnwidth]{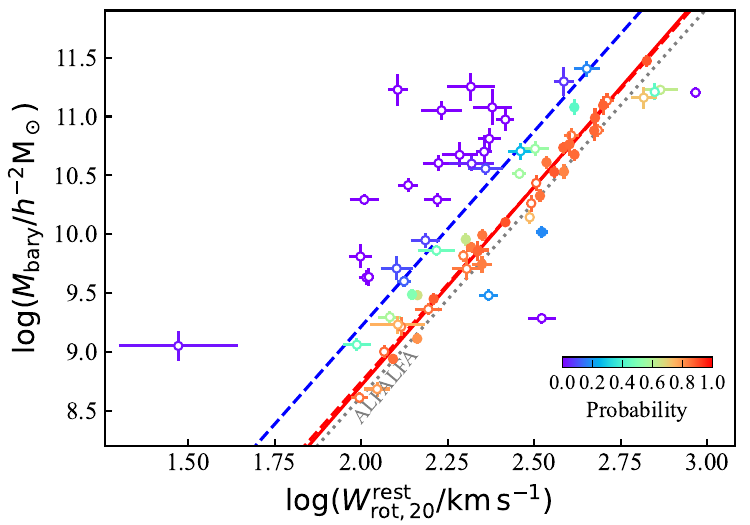}
                \caption{The distribution of the IGS sample in the $M_{\rm bary}$ -- $W_{\rm rot, 20}^{\rm rest}$ plane with the high-quality HQS subset shown with filled dots. The red solid line is the best fit to the HQS using only the ($C_{\rm BTFR}$) component. The best IGS fit is given by the red ($C_{\rm BTFR}$) and blue ($C_{\rm Outlier}$) dashed lines. Note that the red solid and dashed lines almost overlap with each other. The color of dots is coded by probability of a galaxy belonging to the $C_{\rm BTFR}$ component in the IGS sample. The gray dotted line represents the BTFR from ALFALFA based on $W_{\rm peak, 50}$ (converted from $V_{\rm cog,75}$).}\label{Fig_06}
            \end{center}
        \end{figure}

\begin{figure}
            \begin{center}
                \includegraphics[width=0.95\columnwidth]{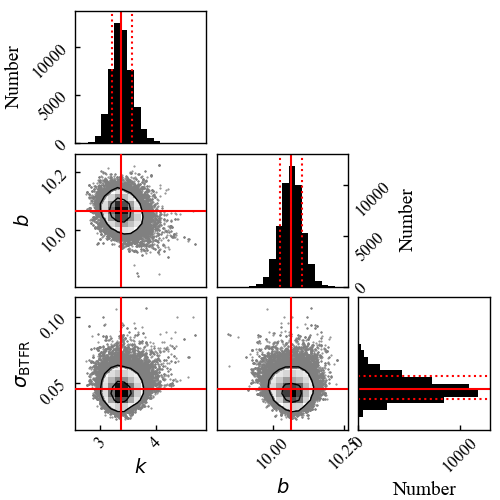}
                \caption{The distribution of data points sampled using the MCMC method on the HQS galaxies in the $M_{\rm bary}$ -- $W_{\rm rot, 20}^{\rm rest}$ plane. The black solid lines indicates the 1- and 2-$\sigma$ contoures (39.3\% and 86.5\% confidence for a 2D Gaussian). The distribution within the 2-$\sigma$ contour is displayed as a gray scale image. The red solid lines indicate the median values, while red dotted lines 1-$\sigma$ 1D uncertainties (15.9\% and 84.1\% percentiles).}\label{Fig_07}
            \end{center}
        \end{figure}

        The 26 \HI\ galaxies in the HQS sample are indicated by filled dots in Figure \ref{Fig_06}. These galaxies have a tight linear relation in the (logarithmic) baryonic mass versus rotation velocity plane. This vindicates the criteria we employed could select a pure sample for BTFR studies. However, the sample is quite small, and has a limited redshift coverage. The distribution has only required a single-component model. Figure \ref{Fig_07} shows the MCMC sampling data points in fitting the model. The best fit values and their uncertainties are listed in the first row of Table \ref{Tab_01}. The best fit line is also shown by red solid line in Figure \ref{Fig_06}.

    \subsection{IGS}

\begin{figure}
            \begin{center}
                \includegraphics[width=0.95\columnwidth]{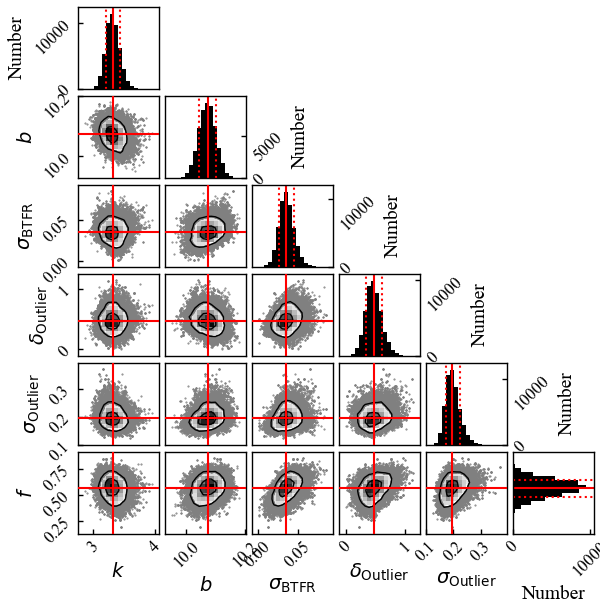}
                \caption{Same as Figure \ref{Fig_07}, but the distribution of IGS sample in $M_{\rm bary}$ -- $W_{\rm rot, 20}^{\rm rest}$ plane fitted by two-component model.}\label{Fig_08}
            \end{center}
        \end{figure}

        The IGS sample has 74 \HI\ galaxies, shown in Figure~\ref{Fig_06}. The distribution has a large scatter in both axes. However, the sample is larger, and has a wider redshift coverage than HQS. We used the best fit line from HQS as a reference to inspect the distribution of IGS sample along the perpendicular direction. It shows two Gaussian components. One is tight and close to the HQS best fit line ($C_{\rm BTFR}$). The other is more extended, with an offset towards lower velocity and higher baryonic mass ($C_{\rm Outlier}$). The amplitude of the former is greater (i.e.\ $f>0.5$ in Equation~\ref{Equ_19}). The two-component model was used to fit the distribution. The MCMC sampling is displayed in Figure \ref{Fig_08}, and the best fit values are given in Table \ref{Tab_01}. The parameters of $C_{\rm BTFR}$ are consistent with that from HQS within 1-$\sigma$ uncertainty, and the results themselves have comparable 1-$\sigma$ uncertainties. Notably, the two-component model can recover the true BTFR in a sample which includes a significant fraction of outliers. We will explore the origin of the lower linewidth and higher mass in the $C_{\rm Outlier}$ component in Section~\ref{Sct_07}. The best fit lines are given by red ($C_{\rm BTFR}$) and blue ($C_{\rm Outlier}$) dashed lines, respectively, in Figure \ref{Fig_06}.
        
        Since the two components overlap, it is hard to separate them. However, we can calculate the probability of each data point belonging to one component using Equation \ref{Equ_20}. The probability is shown by the color bar in Figure \ref{Fig_06}. The data points close to best fit line of $C_{\rm BTFR}$ are more likely to belong to $C_{\rm BTFR}$. Note that the highest probability is $\sim 0.9$, rather than 1.0 since we can not distinguish the two components unambiguously. The probability helps in exploring the origin of the $C_{\rm Outlier}$ component in Section~\ref{Sct_07}.

        \citet{2023ApJ...950...87B} measured the velocity, $V_{\rm cog, 75}$, within which 75\% flux is included in the curve of growth (COG; \citealp{2020ApJ...898..102Y}), as a proxy for rotation velocity in ALFALFA. The baryonic mass was  measured by incorporating stellar mass and 1.33$\times$ the \HI\ mass. They used the same model to fit the BTFR to ALFALFA galaxies, quoting best fit values of $k_{\alpha.R} = 3.30 \pm 0.06$, $b_{\alpha.R}' = 9.900 \pm 0.014$, $\sigma_{\alpha.R, \rm BTFR} = 0.025 \pm 0.004$, $\delta_{\alpha.R, \perp} = -0.07 \pm 0.01$ (offset along direction perpendicular to the best fit line), $\sigma_{\alpha.R, \rm Outlier} = 0.13 \pm 0.01$, and $p_{\alpha.R, \rm Outlier}=0.20  \pm0.03$ (the weighting relative to the main component) using the following equation:
        \begin{equation}
            \log(\frac{M_{\rm bary}}{h_{70}^{-2}{\rm M}_\odot}) = k \times [\log(\frac{V_{\rm cog, 75}}{\rm km\,s^{-1}}) - 2] + b' .
            \label{Equ_22}
        \end{equation}
        We employed the flat correction ($V_{\rm cog, 75} = 0.41 \times W_{\rm peak, 50}$, provided in \citealp{2023ApJ...950...87B}) to convert their results to the parameters used in this paper, propagating the uncertainties assuming the covariances are zero (they do not provide a covariance matrix) using the MC method, giving $b_{\alpha.R}=9.94 \pm 0.01$, $\delta_{\alpha.R, \rm Outlier}=0.24\pm0.03$, and $f_{\alpha.R} = 0.83 \pm 0.02$. The slope is almost identical in both surveys. The value for $\sigma_{\rm BTFR}$ is also consistent within 1-$\sigma$. However, $b$ is smaller in the ALFALFA sample with a significance of $\sim$3-$\sigma$. Differences in definitions of baryonic mass, velocity widths etc.\ could contribute to this offset. The parameters for $C_{\rm Outlier}$ component ($\delta_{\rm Outlier}$ and $\sigma_{\rm Outlier}$) are consistent within 2-$\sigma$, while $f$ is within 3-$\sigma$. It appears that the $C_{\rm Outlier}$ has a similar origin in both surveys.

\section{Discussion}\label{Sct_07}

    The discussion below is based on the results using $W_{\rm m, 20}$ as a measure of rotational velocity. Similar results are derived when using $W_{\rm m, 50}$ (see Appendix \ref{Sct_D}), with some exceptions which are discussed below.

    \subsection{Origin of $C_{\rm Outlier}$}\label{Sct_07_01}

        The HQS is optimal for providing the $C_{\rm BTFR}$ component and providing the best fit for the BTFR. However, we do not know its origin of the outlier component ($C_{\rm Outlier}$). \citet{2023ApJ...950...87B} investigated 80 extreme sources with lowest probability of belonging to $C_{\rm BTFR}$, and found that they all at least have one of the following common characteristics: close companions; uncertain inclinations; inconsistent stellar mass estimates; or inconsistent velocity measures. In this section, we employ a Random Forest (RF) to systematically investigate the contributions to $C_{\rm Outlier}$ component, and their significance.

        The RF is a powerful machine learning algorithm for both regression and classification, making predictions by combining results from a set of decision trees to avoid overfitting. It can handle high dimensional data and gives an importance score for each feature. We use RF as a classifier to inspect the importance of galaxy properties for distinguishing $C_{\rm BTFR}$ from $C_{\rm Outlier}$, including the following parameters: axis ratio ($b/a_{\rm DESI}$) and its uncertainty ($\sigma(b/a_{\rm DESI})$) from the DESI LIS; the difference between DESI LIS and SDSS axis ratios ($\Delta b/a$); the presence of a spectroscopic redshift ($flag_{\rm specz}$); S\'{e}rsic index from DESI LIS photometric models ($n_{\rm sersic}$); \HI\ profile steepness; profile symmetry; number of profile peaks ($N_{\rm peak}$); \HI\ redshift ($z_{\HI}$); mean flux density ($\bar{S_{\nu}}$); $SNR$ of mean flux density; and signal significance ($\ln(P)$). Note that observation limits could also contribute to $C_{\rm Outlier}$ by shifting galaxies towards lower linewidths, e.g., partial detection caused by finite sensitivity, beam smearing caused by finite spatial resolution, and so on. However, quantifying or qualifying these effects for FUDS0 galaxies is beyond the capabilities of current catalog. Galaxies are labeled 1 if the probability of belonging to $C_{\rm BTFR}$ is larger than 0.5, otherwise they are labeled 0. The fraction of $C_{\rm BTFR}$ galaxies is about 86\% given by the best fit parameters of the IGS galaxy distribution. This is also the upper limit of the prediction accuracy.

\begin{figure}
            \begin{center}
                \includegraphics[width=0.95\columnwidth]{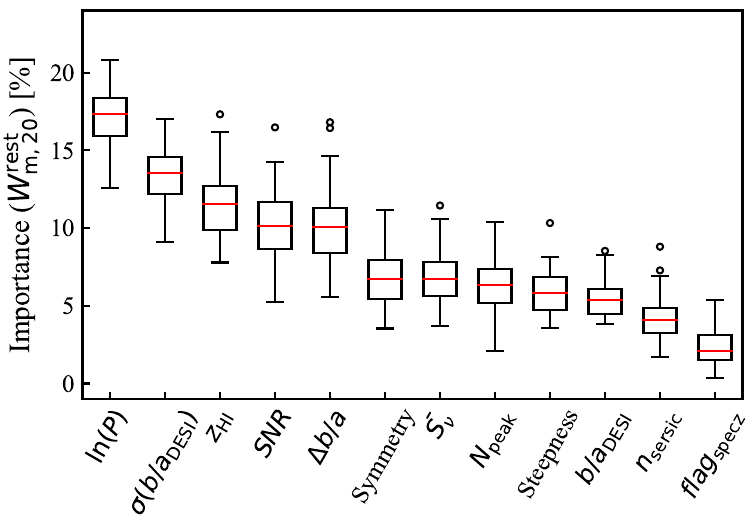}
                \caption{The importance of galaxy properties for studying BTFR in $M_{\rm bary}$ -- $W_{\rm rot, 20}^{\rm rest}$ plane from RF fitting. The boxes represent the interquartile range, and the red lines indicate median values.}\label{Fig_09}
            \end{center}
        \end{figure}

        We utilized the RF implementation \textit{RandomForestClassifier} within the python package \textsc{scikit-learn}\footnote{https://scikit-learn.org/stable/index.html} \citep{scikit-learn}. The RF was optimized using our data with \textit{RandomizedSearchCV} and \textit{GridSearchCV}, which search for the best values of the hyper parameters based on the fitting accuracy from 5-fold cross validation. The optimal values of the hyper parameters are listed in Table \ref{Tab_02}. The default values are adopted for the rest of the hyper parameters. We employed the optimal model to fit a randomly selected 80\% of the data, and evaluated the accuracy using the remaining 20\%. The procedure was repeated 50 times. The average accuracy was 80\%, which is reasonable, and close to our upper limit for prediction accuracy. Finally, the RF gives an importance for each feature, shown in Figure \ref{Fig_09}. We discuss the features in order of importance:

        \begin{table}[!ht]
            \centering
            \caption{The optimized values of hyper parameters in RF.}\label{Tab_02}
            \begin{tabular}{lrrr}
                \toprule
                Parameter & optimal value \\
                \midrule
                \textit{n\_estimators} & 500 \\
                \textit{criterion} & gini \\
                \textit{min\_samples\_split} & 15 \\
                \textit{min\_samples\_leaf} & 7 \\
                \textit{max\_depth} & 10 \\
                \textit{max\_samples} & 0.7 \\
                \textit{max\_features} & 0.2 \\
                \bottomrule
            \end{tabular}
        \end{table}

        \textit{Line strength}: There are three features relating to line strength: $\ln(P)$ (1st), $SNR$ (4th), and $\bar{S_{\nu}}$ (7th). The most important feature is $\ln(P)$, which takes into account the linewidth, amplitude, and noise. The $SNR$ of mean flux density is less important due to the lack of of linewidth information. The difference between $\ln(P)$ and $SNR$ arises from strong narrow lines. Some of these may arise due to partial detection of asymmetric line profiles, thereby underestimating the linewidth. $\bar{S_{\nu}}$ is the least important feature, since it ignores the noise level which changes with frequency and position.

        \textit{Inclination}: Three features are related to inclination: $\sigma(b/a_{\rm DESI})$ (2nd), $\Delta b/a$ (5th), and $b/a_{\rm DESI}$ (10th). $\sigma(b/a_{\rm DESI})$ is the most important since large values result in large and asymmetric uncertainties in rotational velocities after inclination correction, resulting in scatter as well as a systematic shift. The high importance of $\Delta b/a$ derived from different catalogs simply reflects the importance of $\sigma(b/a)$.
        Surprisingly, $b/a$ (or inclination angle itself) has much less impact on studying BTFR. This is due to the inclination cutoff of 40\degr.

        \textit{Distance}: There is only one feature relating to distance, $z_{\HI}$ (3rd). However, many features are indirectly related to distance, including $\ln(P)$, $\sigma(b/a_{\rm DESI})$, $SNR$, and $\bar{S_{\nu}}$. All these features have high importance, therefore contributing to the high importance for $z_{\HI}$. An alternative interpretation is the evolutionary origin of $C_{\rm Outlier}$ caused by evolving kinematic properties of galaxies. However, the higher importance of $\ln(P)$ and $\sigma(b/a_{\rm DESI})$ indicates that the dominant factor is observational effects, rather than evolution. Therefore, nearby galaxies are more suitable for many BTFR studies, but not for studying evolution.
        
        \textit{Confusion}: We use symmetry (6th) and steepness (9th) as indicators of confusion, since smaller symmetry or steepness indirectly represent less confusion. As a result, both features only have intermediate importance, indicating that confusion in FUDS0 appears to have only a small impact on the  BTFR. Note that the main contributor to $M_{\rm bary}$ is $M_*$, rather than $M_\HI$.

        \textit{Line profile}: The number of profile peaks, $N_{\rm peak}$ (8th), is a useful parameter for describing the line profile, especially in separating double-horn profiles from single-peaked profiles. However, it only has intermediate importance for BTFR studies in the $M_{\rm bary}$ -- $W_{\rm rot, 20}^{\rm rest}$ plane. It becomes more important when using $W_{\rm rot, 50}^{\rm rest}$ as a proxy for rotational velocity, where the requirement for double-horn profile helps avoid underestimating $W_{\rm rot, 50}^{\rm rest}$ for extremely asymmetrical line profiles.

        \textit{Morphology}: $n_{\rm sersic}$ (11th) is a proxy for morphology, with low values indicating disk galaxies, and high values indicating the existence of bulge or elliptical galaxies. A possible concern is that the existence of a bulge could lead to an overestimate of $b/a$, therefore an overestimate of rotational velocity. However, the low importance of $n_{\rm sersic}$ suggests that the morphology has little impact on BTFR studies. This is supported by Figure \ref{Fig_06} which shows that only a small fraction of galaxies have higher rotational velocity than the best fit BTFR line.
        
        \textit{Mis-matching}: We cross-matched \HI\ galaxies with their optical counterparts using the set of criteria in our previous work (see \citealp{2025ApJS..278...15X} for details). We use $flag_{\rm specz}$ to represent whether a galaxy has a spectroscopic redshift or not. Galaxies without spectroscopic redshifts may have higher mis-match rates, leading to significant errors in $M_{\rm bary}$. However, $flag_{\rm specz}$ has the least importance among all the features, indicating low impact on BTFR studies, and a low FUDS0 mis-match rate.

        In summary, we infer that low signal significance and large uncertainty of inclination angle are the main contributors to the $C_{\rm Outlier}$ component. Confusion, line profile, inclination angle, and morphology have an intermediate contribution, and mis-match has the least contribution. Line profile is also an important factor, but only if using $W_{\rm m, 50}$ as a proxy for rotational velocity. Our results indicate that $C_{\rm Outlier}$ mainly originates from observational effects. However, we can not rule out the possible contribution from evolution.

    \subsection{Evolution of BTFR}\label{Sct_07_02}

        At low redshift ($z<0.07$), \citet{2024ApJ...976..159D} explored the BTFR using WALLABY galaxies. Consistency with optical spectroscopy from MaNGA was found \citep{2023MNRAS.522.1208A}. \citet{2021MNRAS.508.1195P} did not find any obvious evolution based on 67 galaxies in MIGHTEE-HI survey at redshifts $z \leq 0.081$. At higher redshift, \citet{2015MNRAS.446.3526C} found that 39 HIGHz galaxies in $0.17<z<0.25$ still follow the local BTFR from GASS \citep{2012MNRAS.420.1959C}. \citet{2023MNRAS.519.4279G} showed that the slope of BTFR at $z=0.2$ from BUDHIES is consistent with that from Ursa Major association at $z=0$ \citep{2001ApJ...563..694V}. With the slope fixed, the intercept did not show any obvious evolution.

\begin{figure*}
            \begin{center}
                \includegraphics[width=1.9\columnwidth]{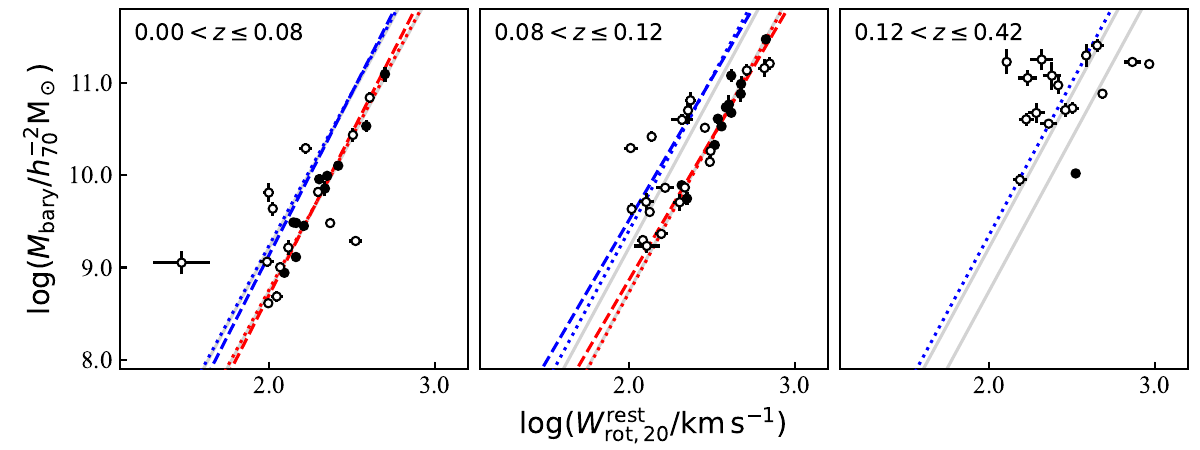}
                \caption{The distribution of IGS galaxies in $M_{\rm bary}$ -- $W_{\rm rot, 20}^{\rm rest}$ plane in three redshift bins. The filled dots indicate the galaxies in the HQS sample. The gray solid lines are the best fits for the full IGS sample. In the first two redshift bins, the dashed lines are the results from fitting both the slope and offset, while the dotted lines are the results from fitting with a fixed slope. In the highest redshift bin, we only show the reliable result for the $C_{\rm Outlier}$ component using one-component fitting with the slope fixed (blue dotted line).}\label{Fig_10}
            \end{center}
        \end{figure*}

        However, the SIMBA hydrodynamic simulation predicts a flattening slope and increasing y-intercept from $z=0$ to 0.5 for star-forming galaxies \citep{2021MNRAS.507.3267G}. The wider redshift coverage of FUDS0 compared with previous work allows a more accurate assessment of evolutionary trends in the BTFR. We therefore divide the IGS sample into three sub-samples at $z=0.08$ and 0.12. The three sub-samples contain 25, 32, and 17 galaxies, respectively. There is a gap around $z=0.12$, where is seriously affected by radio frequency interference (RFI) from globe navigation satellite system (GNSS). Figure \ref{Fig_10} displays the corresponding BTFR for each sub-sample.

        \subsubsection{Slope}

            We examined the galaxy distribution along the direction perpendicular to the best-fit BTFR line from the whole IGS sample. The two components are still clearly observed in the two low-redshift bins. However, the BTFR component vanishes in the highest redshift bin. Both the BTFR and outlier components have similar positions and scatter in different redshift bins. Hence, we also employed the two-component model to fit the galaxy distribution in the three redshift bins. The best-fit values are listed in Table \ref{Tab_01}. As we mentioned above, there is only one component ($C_{\rm Outlier}$) left in the highest redshift bin, and the galaxies are concentrated in narrow ranges of linewidth and baryonic mass. It is hard to measure an accurate slope for $C_{\rm BTFR}$ in this bin. Hence, Figure \ref{Fig_10} only shows the results for the two low-redshift bins.

            The resultant slopes of $C_{\rm BTFR}$ are consistent with each other within 1-$\sigma$ uncertainty in the two low-redshift sub-samples. There is also consistency with the local BTFR from ALFALFA \citep{2023ApJ...950...87B}. These results indicate no significant evolution of the BTFR slope. Although $C_{\rm BTFR}$ vanishes in the highest redshift bin, the shared slope derived from $C_{\rm Outlier}$ also implies the consistent slope of $C_{\rm BTFR}$, which aligns with the conclusion from the two low-redshift bins.

        \subsubsection{Zero point}

            We also inspect the change of zero point in different redshift bins. Considering no significant evolution of slope is detected, the fit was made using the two-component model with the slope fixed using the value from the full IGS sample. The best fit values are tabulated in Table \ref{Tab_01}. The results again show a consistent zero point of $C_{\rm BTFR}$ ($b$) within 1-$\sigma$ uncertainty in the two low redshift bins -- i.e.\ no significant evolution.
            
            As mentioned already, there is only one component ($C_{\rm Outlier}$) in the highest redshift bin. The best-fit results do not provide useful information about $C_{\rm BTFR}$. However, it is possible to derive $b$ from $C_{\rm Outlier}$ assuming a fixed offset $\delta_{\rm Outlier}$ to indirectly explore the evolution of the zero point. We therefore fit the one-component model with fixed slope $k$, using $\delta_{\rm Outlier}$ from the full IGS sample. For comparison, the zero point for $C_{\rm Outlier}$ is also computed as the sum of $b$ and $\delta_{\rm Outlier}$. The best-fit line for $C_{\rm Outlier}$ is shown in the highest-redshift panel of Figure \ref{Fig_10}, while the best-fit results are listed in Table \ref{Tab_01}. The indirectly derived zero point of $C_{\rm BTFR}$ remains consistent with the conclusion of no significant evolution.

            Although the observational effects dominate the origin of $C_{\rm Outlier}$, we still cannot rule out the possibility of a transformation from $C_{\rm BTFR}$ to $C_{\rm Outlier}$ caused by evolution (as mentioned in Section \ref{Sct_07_01}). Our non-detection of evolution could be due to a combination of observational and evolutionary effects. Genuine evolution might be masked by the observational effects. However, distinguishing them is beyond the capability of the current data.

\section{Summary}\label{Sct_08}

    FUDS0 provides \HI\ pencil-beam detections of 128 galaxies over the redshift range $0<z<0.4$. In this paper, we study the baryonic Tully-Fisher relation (BTFR) over the same redshift range. Stellar masses have been computed from fitting the Spectral Energy Distribution (SED) using UV, optical and IR data. Our `inclined galaxy sample' (IGS) is restricted to those with inclination angles larger than 40\degr\ and baryonic masses larger than $10^{8.3} h^{-2} {\rm M}_\odot$, of which there are 74 galaxies. More stringent criteria are used to define a `high quality sample' (HQS) of 24 galaxies. Our main results are given below:

    \begin{enumerate}
    
        \item The HQS sample shows tight relation in $M_{\rm bary}$ -- $W_{\rm rot, 20}^{\rm rest}$ plane. A single-component model ($y=k \times (x-2.4) + b$) was fit to the data, resulting in the best-fit parameters: slope $k=3.38_{-0.17}^{+0.20}$, zero point $b=10.07_{-0.04}^{+0.04}$, and intrinsic scatter $\sigma_{\rm BTFR} = 0.045_{-0.007}^{+0.010}$.

        \item The IGS sample has a larger scatter than HQS, with an extra outlier component ($C_{\rm Outlier}$) located at upper-left to the BTFR component ($C_{\rm BTFR}$). A two-component model was use to fit this distribution, yielding similar parameters, $k = 3.32_{-0.11}^{+0.12}$ and $b = 10.07_{-0.03}^{+0.03}$, and $\sigma_{\rm BTFR} = 0.036_{-0.009}^{+0.010}$, thereby demonstrating the robustness of recovery of the BTFR using samples with significant outliers.

        \item The origin of the outlier component ($C_{\rm Outlier}$) in the IGS sample was investigated using a random forest algorithm. We found that low signal significance and inaccurate inclinations are the main contributing components to the outliers, indicating that $C_{\rm Outlier}$ originates mainly from observational effects.

        \item The IGS sample was split into three sub-samples with boundaries at redshifts $z=0.08$ and $0.12$. The BTFR was directly calculated in the two low-redshift bins with a two-component model. Both slopes and zero points are consistent within the 1-$\sigma$ uncertainty, indicating no significant evolution. Due to the lack of $C_{\rm BTFR}$, the BTFR is indirectly inferred in the highest-redshift bin, showing consistency with the results from the low-redshift bins and aligning with the conclusion. Given the possible origin of $C_{\rm Outlier}$, the evolution of the BTFR might be masked by the observational effects.

    \end{enumerate}

    When complete, FUDS will provide six times the number of \HI\ galaxies to further explore BTFR evolution for $z<0.42$, complementing interferometric surveys such as DINGO ($z<0.43$), MIGHTEE ($z<0.6$) and LADUMA ($z<1.3$).

%% Also note that the akcnowlodgment environment does not support long amounts of text. If you have a lot of people and institutions to acknowledge, do not use this command. Instead, create a new \section{Acknowledgments}.
\begin{acknowledgments}

    We thank the anonymous referee for constructive suggestions on improving the integrity and logic of the paper. This work made use of the data from FAST (Five-hundred-meter Aperture Spherical radio Telescope) (\url{https://cstr.cn/31116.02.FAST}). FAST is a Chinese national mega-science facility, operated by National Astronomical Observatories, Chinese Academy of Sciences. The work is supported by the National Key R\&D Program of China under grant number 2018YFA0404703, the Science and Technology Innovation Program of Hunan Province under grant number 2024JC0001, 2025RC4002, the Investigation of Technological Infrastructure Resources under grant number 2023FY101101, and the FAST Collaboration. Parts of this research were supported by the Australian Research Council Centre of Excellence for All Sky Astrophysics in 3 Dimensions (ASTRO 3D), through project number CE170100013.
    
\end{acknowledgments}


\vspace{5mm}

\facilities{FAST}


\software{Matplotlib \citep{Hunter:2007}, SciPy \citep{2020SciPy-NMeth}, NumPy \citep{harris2020array}, AstroPy \citep{2013A&A...558A..33A, 2018AJ....156..123A, 2022ApJ...935..167A}, scikit-learn \citep{scikit-learn}, EMCEE \citep{2013PASP..125..306F}}



\appendix

\section{Redshifts}\label{Sct_A}

    \subsection{Solar motion}

        By analysing the data from Cosmic Background Explorer (COBE), \citet{1996ApJ...473..576F} derived the heliocentric motion\footnote{The difference between heliocentric and barycentric velocities is $\sim 20$ m s$^{-1}$ at maximum, mainly caused by the orbital motion of Jupiter with period of $\sim 12$ years. The difference is only $\sim 1.2\%$. Hence, we just ignore the difference, and use $V^{\rm bary} = V^{\rm hel}$.}, $V_{\rm hel}^{\rm cmb}=371.0$ km s$^{-1}$ toward $(l_{\rm hel}, b_{\rm hel}) = (264.\!\!^\circ14, 48.\!\!^\circ26)$, relative to Cosmic Microwave Background (CMB). The following equation\footnote{\url{https://ned.ipac.caltech.edu/Documents/Guides/Calculators}} was used to calculate the velocity of barycenter in CMB frame projected to the target galaxy's direction $(l, b)$.
        \begin{equation}
            \begin{split}
                V_{\rm bary, proj}^{\rm cmb} = & V_{\rm hel}^{\rm cmb} [\cos(b)\cos(b_{\rm hel})cos(l-l_{\rm hel})\\
                &  + \sin(b)\sin(b_{\rm hel})]
            \end{split}\label{Equ_A1}
        \end{equation}
        Then the redshift caused by the heliocentric motion could be derived by relativistic equation.
        \begin{equation}
            z_{\rm hel, proj}^{\rm cmb} = \sqrt{\frac{c+V_{\rm hel, proj}^{\rm cmb}}{c-V_{\rm hel, proj}^{\rm cmb}}} - 1 \label{Equ_A2}
        \end{equation}
        in which $c$ is the speed of light in vacuum.

    \subsection{Galaxy motion}

        Taking into account the correction at high redshift, we derived the velocity modified by cosmological curvature term by the following equation \citep{2016AJ....152...50T}.
        \begin{equation}
            \begin{split}
                V_{\rm mod}^{\rm [frame]} = & cz^{\rm [frame]}[1 + \frac{1}{2}(1 - q_{\rm 0, cf3})z^{\rm [frame]}\\
                &- \frac{1}{6}(1 - q_0 - 3q_0^2 + j_0)(z^{\rm [frame]})^2] \label{Equ_A3}
            \end{split}
        \end{equation}
        where $q_0 = \frac{1}{2} (\Omega_{\rm M} - 2\Omega_{\Lambda})$, $j_0 = 1$ in a flat universe. Therefore, the projected peculiar velocity can be computed using the following equation \citep{2016AJ....152...50T}:
        \begin{equation}
            \begin{split}
                V_{\rm pec, proj}^{\rm [frame]} = & (V_{\rm mod}^{\rm [frame]} - H_0 D_{\rm lum}^{\rm [frame]})\\
                &/(1 + H_0 D_{\rm lum}^{\rm [frame]}/c) .
            \end{split}\label{Equ_A4}
        \end{equation}
        The frame is either GSR in NAM or LS in CF3. Accordingly, we adopted the cosmological parameters ($\Omega_{\rm M, cf3}$, $\Omega_{\rm \Lambda, cf3}$) and Hubble constant ($H_{\rm 0, cf3}$) from Cosmicflows-3. Combining Equation \ref{Equ_04} (or \ref{Equ_05}) and \ref{Equ_A1}, we can convert the peculiar velocity from GSR (or LS) to the CMB frame. The redshift caused by peculiar velocity is then given by relativistic equation.
        \begin{equation}
            z_{\rm pec, proj}^{\rm cmb} = \sqrt{\frac{c+V_{\rm pec, proj}^{\rm cmb}}{c-V_{\rm pec, proj}^{\rm cmb}}} - 1 . \label{Equ_A5}
        \end{equation}

    \subsection{Cosmological redshift}

        The cosmological redshift is then computed using the following equation to remove solar and Galactic peculiar motions:
        \begin{equation}
            z_{\rm cos}^{\rm cmb} = (1 + z_{\rm obs}^{\rm hel})(1 + z_{\rm hel, proj}^{\rm cmb})/(1 + z_{\rm pec, proj}^{\rm cmb})-1 . \label{Equ_A6}
        \end{equation}

\section{Distances}\label{Sct_B}

    The comoving distance, $D_{\rm com}$, was computed from $z_{\rm cos}^{\rm cmb}$ (see Appendix \ref{Sct_A}) with the cosmological parameters ($\Omega_{\rm M}$, $\Omega_{\Lambda}$) and Hubble constant ($H_0$) used in this paper. The luminosity distance in heliocentric frame can be derived using the following equation:
    \begin{equation}
        \begin{split}
            D_{\rm lum}^{\rm hel} & = (1 + z_{\rm cmb, proj}^{\rm hel})(1 + z_{\rm cos}^{\rm cmb})(1 + z_{\rm pec, proj}^{\rm cmb})^2 D_{\rm com} \\
            & = \frac{1}{1 + z_{\rm hel, proj}^{\rm cmb}}(1 + z_{\rm cos}^{\rm cmb})(1 + z_{\rm pec, proj}^{\rm cmb})^2 D_{\rm com} . \label{Equ_B7}
        \end{split}
    \end{equation}
    in which the square includes both Doppler and Beaming effects from special relativity. Note that $1 + z_{\rm cmb, proj}^{\rm hel} = 1/(1 + z_{\rm hel, proj}^{\rm cmb})$.

\section{Simulation}\label{Sct_C}

    We performed a simulation to mock the FUDS0 catalog. The target sky is centered at the center of the FUDS0 field ($\alpha=$124\fdg3, $\delta=$22\fdg18) with an extension of 1\degr\ in both the $R.A.$ and $Dec$ directions. The redshift coverage is from 0 to 0.42. We do not take into account the large-scale structure in the volume for simplicity. The stellar mass ($M_*$) is sampled from the stellar mass function (SMF) with parameters from \citet{2022MNRAS.513..439D}. Then, we computed the \HI\ mass ($M_\HI$) based on the scaling relation derived in \citet{2023MNRAS.525..256P}. We employed the same method described in Section \ref{Sct_04_02_03} to calculate the molecular mass ($M_{\rm Mol}$). The baryonic mass ($M_{\rm bary}$) is then derived by Equation \ref{Equ_12}. We employed the BTFR to calculate the rotational velocity ($W_{\rm rot}^{\rm rest}$) with parameters of $k=3.3$, $b=10.0$, $\sigma_{\rm BTFR}=0.04$ at a reference velocity of $\log(W_{\rm rot}^{\rm rest})=2.4$ to mimic a one-component relation. To mimic a two-component relation, we introduced another line with same slope, but with $b_{\rm Outlier}=10.5$ (according to $\delta_{\rm Outlier} = b_{\rm Outlier} - b = 0.5$), and $\sigma_{\rm Outlier} = 0.2$. $f=0.6$ is adopted as the mixture factor. The inclination angle ($\theta_{\rm inc}$) is assigned to galaxies by randomly pointing the rotational axes in space, before rescaling by Equation \ref{Equ_09} with minima of $q_0=0.2$ to derive the axial ratio ($b/a$). Then the projected linewidth ($W_{\rm rot, proj}^{\rm rest}$) is computed by projecting $W_{\rm rot}^{\rm rest}$ to the line-of-sight. The turbulent velocity is described as a function of redshift by $W_{\rm tur} = 30 + mz$, where $m=11$ from \citet{2019ApJ...880...48U}. Equation \ref{Equ_06} is adopted to derive the final linewidth ($W_{\rm rot, proj}^{\rm obs}$) by mixing rotational and turbulent velocities.

    The uncertainties of the observables are estimated by scaling relations in the following spaces based on the FUDS0 catalog:
    \begin{itemize}
    
        \item $\log(\frac{\sigma_{S_{\rm int}}}{\rm Jy\,Hz}) - \log({\frac{\rm RMS}{\rm Jy}}\sqrt{\frac{W_{\rm rot, proj}^{\rm obs}}{\rm Hz} \frac{\rm Hz}{W_{\rm channel}}})$
        
        \item $\log(\frac{\sigma_{W_{\rm rot, proj}^{\rm obs}}}{\rm Hz}) - \log(\frac{S_{\rm int}}{\rm Jy} \frac{\rm Jy}{\rm RMS} \sqrt{\frac{W_{\rm channel}}{\rm Hz} \frac{\rm Hz}{W_{\rm rot, proj}^{\rm obs}}})$
        
        \item $\log(\sigma_{b/a}) - \log(\frac{M_*}{h_{70}^{-2} {\rm M_\odot}} (\frac{h_{70}^{-1}{\rm Mpc}}{D_{\rm Lum}})^2)$

        \item $\log(\frac{\sigma_{M_*}}{h_{70}^{-2} {\rm M_\odot}}) - \log(\frac{M_*}{h_{70}^{-2} {\rm M_\odot}})$
        
    \end{itemize}
    where RMS is the local noise in the FUDS0 space-space-frequency cube. To mimic the deviation of observables from ground-truth values, we resample $S_{\rm int}$, $W_{\rm rot, proj}^{\rm obs}$, $b/a$, and $M_*$ from a Gaussian distribution with means and uncertainties derived above. Then the median and uncertainties of $M_\HI$, $M_{\rm Mol}$, $M_{\rm bary}$, $W_{\rm rot, proj}^{\rm rest}$, $\theta_{\rm inc}$, and $W_{\rm rot}^{\rm rest}$ are computed by MC sampling as observables.
    
    The completeness of the FUDS0 catalog \citep{2024ApJS..274...18X} is employed to select detectable galaxies in the simulation. The criteria for IGS are also adopted to select suitable galaxies for the BTFR study. We take into account the redshift distributions of IGS and HQS. The same numbers of galaxies are randomly selected in each narrow redshift bin to generate mock IGS and HQS.

\section{Results in the $M_{\rm bary}$ -- $W_{\rm rot, 50}^{\rm rest}$ plane}\label{Sct_D}

    \subsection{HQS}

\begin{figure}
            \begin{center}
                \includegraphics[width=0.95\columnwidth]{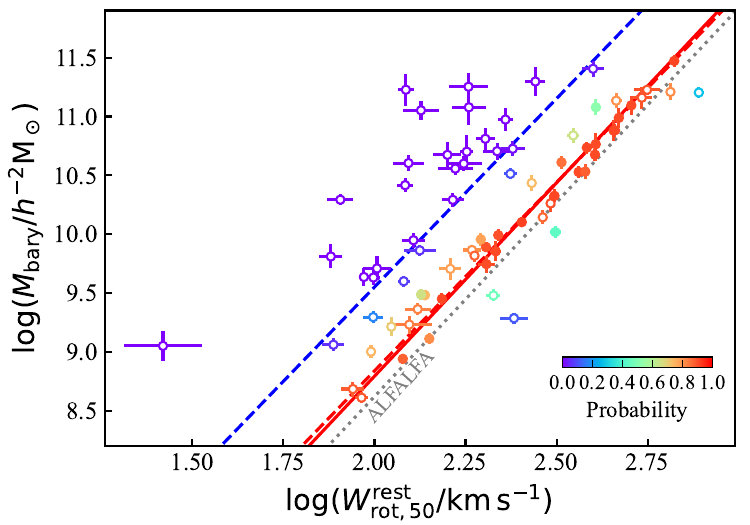}
                \caption{Same as Figure \ref{Fig_06}, but in the $M_{\rm bary}$ -- $W_{\rm rot, 50}^{\rm rest}$ plane.}\label{Fig_11}
            \end{center}
        \end{figure}

\begin{figure}
            \begin{center}
                \includegraphics[width=0.95\columnwidth]{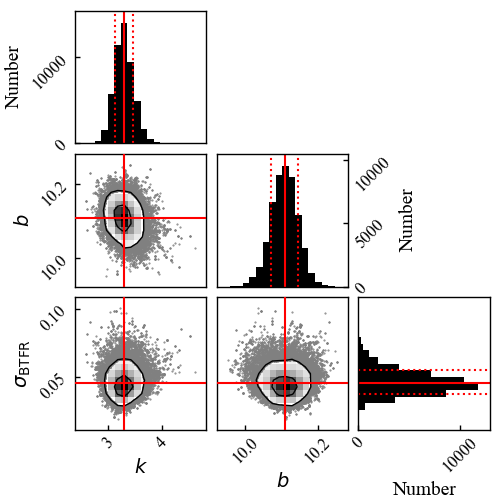}
                \caption{Same as Figure \ref{Fig_07}, but fitting in the $M_{\rm bary}$ -- $W_{\rm rot, 50}^{\rm rest}$ plane.}\label{Fig_12}
            \end{center}
        \end{figure}

        We use a one-component model to fit the distribution of HQS galaxies in the $M_{\rm bary}$ -- $W_{\rm rot, 50}^{\rm rest}$ plane. In Figure \ref{Fig_11}, the HQS galaxies are shown as filled dots. The best fit line is given by the red solid line. The sampled MCMC data points are displayed in Figure \ref{Fig_12}. The best fit values of BTFR parameters are listed in the first row of Table \ref{Tab_03}.

        \begin{table*}[!ht]
            \centering
            \caption{Same as Table \ref{Tab_01}, but the best fit values in $M_{\rm bary}$ -- $W_{\rm rot, 50}$ plane.}\label{Tab_03}
            \begin{tabular}{lrrrrrrrrr}
                \toprule
                    &  & & \multicolumn{6}{c}{Parameters} \\
                \cmidrule(lr){4-10}
                Sample & $z$ & $N_{\rm smple}$ & $k$ & $b$ & $\sigma_{\rm BTFR}$ & $\delta_{\rm Outlier}$ & $\sigma_{\rm Outlier}$ & $f$ & $b+\delta_{\rm Outlier}$\\
                  & & & [2, 5] & [2, 15] & [0.001, 0.25] & [0, 2] & [0.001, 0.5] & [0.01, 0.99] & - \\
                \midrule

                HQS & $0.00 - 0.42$ &24 & $3.28_{-0.17}^{+0.18}$ & $10.11_{-0.04}^{+0.04}$ & $0.045_{-0.008}^{+0.010}$ & - & - & - & - \\
                \cmidrule(lr){1-10}
                \multirow{8}{*}{IGS} & $0.00 - 0.42$ & 74 & $3.20_{-0.11}^{+0.12}$ & $10.12_{-0.03}^{+0.03}$ & $0.041_{-0.008}^{+0.011}$ & $0.71_{-0.15}^{+0.17}$ & $0.20_{-0.03}^{+0.03}$ & $0.58_{-0.08}^{+0.08}$ & $10.83_{-0.15}^{+0.17}$ \\
                \cmidrule(lr){2-10}
                & \multirow{2}{*}{$0.00 - 0.08$} & \multirow{2}{*}{25} & $3.40_{-0.32}^{+0.41}$ & $10.20_{-0.09}^{+0.10}$ & $0.068_{-0.024}^{+0.033}$ & $0.71_{-0.43}^{+0.61}$ & $0.26_{-0.08}^{+0.12}$ & $0.71_{-0.18}^{+0.14}$ & $10.89_{-0.42}^{+0.61}$\\
                &  & & $3.20$ [fix] & $10.17_{-0.06}^{+0.07}$ & $0.062_{-0.020}^{+0.032}$ & $0.62_{-0.37}^{+0.53}$ & $0.26_{-0.08}^{+0.12}$ & $0.70_{-0.16}^{+0.14}$ & $10.79_{-0.36}^{+0.53}$ \\
                \cmidrule(lr){2-10}
                & \multirow{2}{*}{$0.08 - 0.12$} & \multirow{2}{*}{32} & $2.96_{-0.28}^{+0.27}$ & $10.16_{-0.07}^{+0.08}$ & $0.040_{-0.017}^{+0.025}$ & $0.70_{-0.21}^{+0.21}$ & $0.15_{-0.04}^{+0.05}$ & $0.61_{-0.15}^{+0.15}$ & $10.85_{-0.21}^{+0.23}$\\
                &  & & $3.20$ [fix] & $10.11_{-0.04}^{+0.05}$ & $0.032_{-0.012}^{+0.018}$ & $0.74_{-0.19}^{+0.18}$ & $0.15_{-0.03}^{+0.04}$ & $0.58_{-0.13}^{+0.12}$ & $10.85_{-0.19}^{+0.20}$ \\
                \cmidrule(lr){2-10}
                & \multirow{3}{*}{$0.12 - 0.42$} & \multirow{3}{*}{17} & $4.00_{-0.94}^{+0.70}$ & $10.50_{-0.82}^{+0.53}$ & $0.160_{-0.086}^{+0.056}$ & $1.05_{-0.75}^{+0.60}$ & $0.24_{-0.07}^{+0.11}$ & $0.35_{-0.23}^{+0.46}$ & $11.31_{-0.31}^{+0.54}$\\
                &  & & $3.20$ [fix] & $10.35_{-0.63}^{+0.61}$ & $0.130_{-0.078}^{+0.077}$ & $1.01_{-0.79}^{+0.50}$ & $0.23_{-0.07}^{+0.09}$ & $0.26_{-0.17}^{+0.42}$ & $11.19_{-0.24}^{+0.29}$ \\
                 &  & & $3.20$ [fix] & $10.26_{-0.19}^{+0.20}$ & - & $0.71$ [fix] & $0.23_{-0.04}^{+0.05}$ & - & $10.97_{-0.19}^{+0.20}$ \\
                
                \bottomrule
            \end{tabular}
        \end{table*}

    \subsection{IGS}

\begin{figure}
            \begin{center}
                \includegraphics[width=0.95\columnwidth]{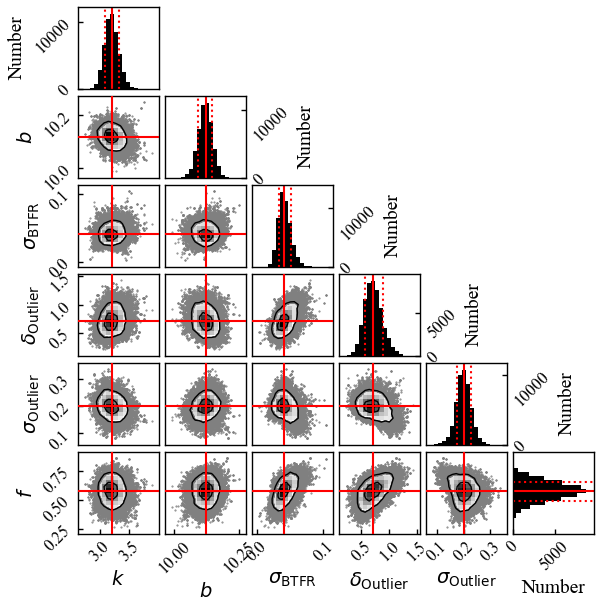}
                \caption{Same as Figure \ref{Fig_08}, but fitting results in the $M_{\rm bary}$ -- $W_{\rm rot, 50}^{\rm rest}$ plane.}\label{Fig_13}
            \end{center}
        \end{figure}

        The two-component model was employed to fit the distribution of IGS galaxies in $M_{\rm bary}$ -- $W_{\rm rot, 50}^{\rm rest}$ plane. The IGS galaxies are shown in Figure \ref{Fig_11} with the best fit lines given by red ($C_{\rm BTFR}$) and blue ($C_{\rm Outlier}$) dashed lines. The best fit values are listed in Table \ref{Tab_03}. We also show the data points from MCMC sampling in Figure \ref{Fig_13}. The best fit values of BTFR line are consistent with that from HQS within the 1-$\sigma$ uncertainty.

    \subsection{Importance}

\begin{figure}
            \begin{center}
                \includegraphics[width=0.95\columnwidth]{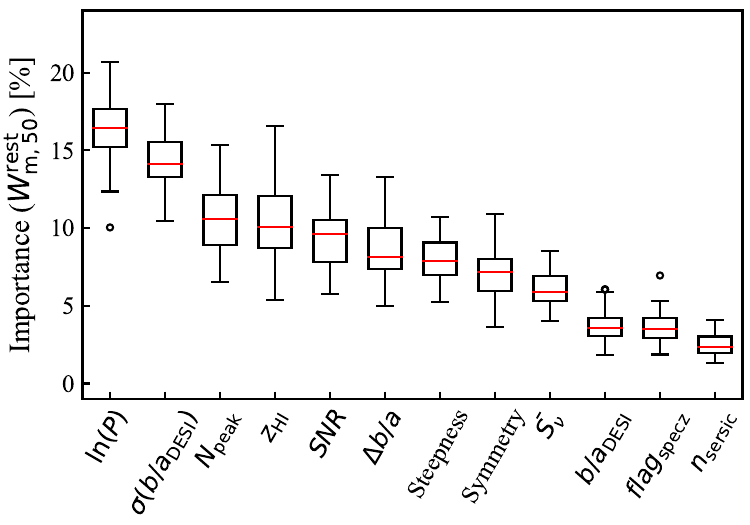}
                \caption{Same as Figure \ref{Fig_09}, but in the $M_{\rm bary}$ -- $W_{\rm rot, 50}^{\rm rest}$ plane.}\label{Fig_14}
            \end{center}
        \end{figure}

        Based on the probability of each data point belonging to $C_{\rm BTFR}$ component, the RF was employed to explore the importance of galaxy properties for BTFR studies in the $M_{\rm bary}$ -- $W_{\rm rot, 50}^{\rm rest}$ plane, as displayed in Figure \ref{Fig_14}. It is similar to that already derived in $M_{\rm bary}$ -- $W_{\rm rot, 20}^{\rm rest}$ plane. The signal significance and uncertainty in inclination angle are still the dominant features for BTFR studies. However, the number of peaks ($N_{\rm peak}$) has much higher importance than for the $M_{\rm bary}$ -- $W_{\rm rot, 20}^{\rm rest}$ plane. This indicates that $N_{\rm peak}$ is a key feature for BTFR studies, which effectively avoids underestimation of rotational velocity in galaxies with extremely asymmetric line profiles.

    \subsection{Evolution}

\begin{figure*}
            \begin{center}
                \includegraphics[width=1.9\columnwidth]{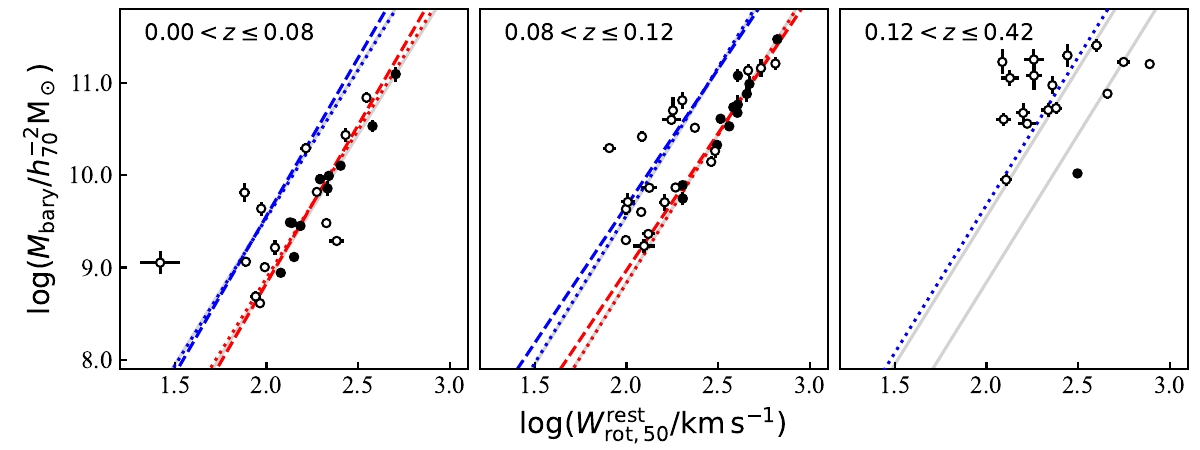}
                \caption{Same as Figure \ref{Fig_10}, but in the $M_{\rm bary}$ -- $W_{\rm rot, 50}^{\rm rest}$ plane.}\label{Fig_15}
            \end{center}
        \end{figure*}

        The IGS sample was split into three sub-samples at redshifts $z=0.08$ and 0.12. In Figure \ref{Fig_15}, we show the distribution of IGS galaxies in three redshift bins in the $M_{\rm bary}$ -- $W_{\rm rot, 50}^{\rm rest}$ plane. The two-component model was employed to fit the galaxy distribution in each sub-sample to explore the evolution of the slope. The best-fit values are given in Table \ref{Tab_03}. In the two low-redshift bins, the best-fit slope is consistent with that from the full IGS sample within the 1-$\sigma$ uncertainty. The result indicates no significant evolution of the BTFR slope. Since there is only one component ($C_{\rm Outlier}$) in the highest redshift bin, the parameters for the $C_{\rm BTFR}$ component, e.g., $k$, $b$, and $\sigma_{\rm BTFR}$, have large 1-$\sigma$ uncertainties. The shared slope from $C_{\rm Outlier}$ is also consistent with that in the two low-redshift bins, aligning with the conclusion. Given the large uncertainties in the highest redshift bin, we only display the best-fit lines in the two low-redshift bins in Figure \ref{Fig_15} using dashed lines. The colors indicate different components (red for $C_{\rm BTFR}$ and blue for $C_{\rm Outlier}$).
        
        In order to study the evolution of the zero point, we performed a two-component fit with the slope fixed to that derived for the full IGS sample considering no significant evolution for the BTFR slope. The best-fit results are given in Table \ref{Tab_03}. The zero point is consistent with that of the full IGS sample within 1-$\sigma$ in the two low-redshift bins, indicating no significant evolution. As mentioned, there is only one component ($C_{\rm Outlier}$) in the highest redshift bin, so the best-fit results from the two-component model are not appropriate for the $C_{\rm BTFR}$ component. Hence, we use the one-component model to fit the distribution with the slope fixed. The zero point of the BTFR is indirectly inferred by using $\delta_{\rm Outlier}$ from the full IGS sample. The best-fit values are also given in Table \ref{Tab_03}. Both $b$ and $b+\delta_{\rm Outlier}$ are consistent within 1-$\sigma$ with the results from the two low-redshift bins. The result aligns with the conclusion on the zero point from the two low-redshift bins. In Figure \ref{Fig_15}, we display the results from the one-component model in the highest-redshift bin.

\bibliography{References}{}
\bibliographystyle{aasjournal}



\end{document}